\documentclass[conference]{IEEEtran}
\usepackage{times}  %Required
\usepackage{helvet}  %Required
\usepackage{courier}  %Required
\usepackage{url}  %Required
\usepackage{graphicx}  %Required
\usepackage{comment}
\usepackage{amsmath}
\usepackage{algorithm}
\usepackage[noend]{algpseudocode}
\usepackage{comment}
\usepackage{amssymb}
\usepackage{subfig}
\usepackage{graphicx} % more modern
\usepackage{color}
\usepackage{hyperref}

\ifCLASSINFOpdf
\else
\fi
\IEEEoverridecommandlockouts

\begin{document}
\title{Attack and Defense of Dynamic Analysis-Based,\\
Adversarial Neural Malware Classification Models}

\author{\IEEEauthorblockN{Jack W. Stokes\footnotemark{*}\thanks{* Jack Stokes and De Wang made equal contributions to this work.}}
\IEEEauthorblockA{Microsoft Resarch\\
Redmond, WA 98052}
\and
\IEEEauthorblockN{De Wang\footnotemark{*}}
\IEEEauthorblockA{University of Texas at Arlington\\
Arlington, TX 76019}
\and
\IEEEauthorblockN{Mady Marinescu, Marc Marino, Brian Bussone}
\IEEEauthorblockA{Microsoft Corporation\\
Redmond, WA 98052}
}


%Telephone: (800) 555--1212, Fax: (888) 555--1212}


\maketitle
\begin{abstract}
Recently researchers have proposed using deep learning-based systems for malware detection. Unfortunately, all deep learning classification systems are vulnerable to adversarial attacks where miscreants can avoid detection by the classification algorithm with very few perturbations of the input data. Previous work has studied adversarial attacks against static analysis-based malware classifiers which only classify the content of the unknown file without execution. However, since the majority of malware is either packed or encrypted, malware classification based on static analysis often fails to detect these types of files. To overcome this limitation, anti-malware companies typically perform dynamic analysis by emulating each file in the anti-malware engine or performing in-depth scanning in a virtual machine. These strategies allow the analysis of the malware after unpacking or decryption. In this work, we study different strategies of crafting adversarial samples for dynamic analysis. These strategies operate on sparse, binary inputs in contrast to continuous inputs such as pixels in images. We then study the effects of two, previously proposed defensive mechanisms against crafted adversarial samples including the distillation and ensemble defenses. We also propose and evaluate the weight decay defense.
Experiments show that with these three defensive strategies, the number of successfully crafted adversarial samples is reduced compared to a standard baseline system without any defenses.
In particular, the ensemble defense is the most resilient to adversarial attacks.
Importantly, none of the defenses significantly reduce the classification accuracy for detecting malware.
Finally, we demonstrate that while adding additional hidden layers to neural models does not significantly improve the malware classification accuracy, it does significantly increase the classifier's robustness to
adversarial attacks.
\end{abstract} 

% \fi
\IEEEpeerreviewmaketitle

\section{Introduction}
Cybersecurity is an important emerging application in artificial intelligence. As commercial and open source software authors improve the security of their applications, and organizations
deploy advanced threat detection systems to harden their defenses, attackers will be forced to employ more sophisticated attacks in order to infect a computer or penetrate
an organization's network. One of the primary computer security defenses continues to be commercial anti-malware products.
A number of researchers~\cite{dahl2013large,saxe2015,pascanu2015malware,Huang2016,BenMalware,Kolosnjaji} have proposed the use of deep learning for malware classification as a key component of next generation anti-malware systems.

Recently, researchers have also started to study the attacks and defenses of machine learning-based classification systems, and this area is commonly known as adversarial learning.
In an adversarial learning-based attack, miscreants intentionally craft malicious samples which are designed to confuse (\emph{i.e.}, fool) a deployed machine learning model.
An adversarial sample is one whose input data is altered
%/perturbed from original the input data
in such a way that the perturbation does not change its ground truth label, but the altered sample is misclassified by a trained machine learning model. In some cases, such as images~\cite{papernot2016practical},
 % or speech~\cite{},
the goal is to alter these samples in such a way that they are not perceived by humans to be intentionally corrupted. To be more specific, by perturbing a tiny fraction of the raw input vector features (e.g., pixels) or adding noise with a very small magnitude compared to the original input vector~\cite{goodfellow2014explaining}, the crafted sample will be misclassified as
belonging to a different class. In some cases, the attacker decides to target the mispredicted class to be any desired class. It is a phenomenon that has appeared in some of the deep learning literature~\cite{goodfellow2014explaining,nguyen2015deep}, but it also exists in shallow linear models~\cite{weilinxu2016evading}.

While many authors have focused on adversarial learning-based attacks, only a few defenses have been proposed. Goodfellow, \emph{et al.},~\cite{goodfellow2014explaining} proposed training with adversarial samples. In 2015, Papernot, \emph{et al.},~\cite{papernot2015distillation} proposed the distillation defense for adversarial learning. More recently, several authors have proposed an ensemble defense~\cite{Kantchelian2016,Tramer2016,Feng2016,Tramer2017} for adversarial samples. Xu, \emph{et al.},~\cite{Xu2017}  proposed a feature squeezing system to detect potential adversarial samples by measuring the difference between the original model and a new model where unnecessary input features have been removed.

Most of the previous research in adversarial learning has typically focused on non-adversarial datasets such as
images ~\cite{papernot2016practical,goodfellow2014explaining}. Malware classification, on the other hand, is arguably one of the most
adversarial environments. To date, relatively few studies have investigated adversarial learning in the field of malware classification. Several papers have focused on the attack side. Hu, \emph{et al.},~\cite{weilinxu2016evading} study adversarial learning
in the context of linear classifiers which are designed to detect malicious PDF (\emph{i.e.}, Adobe Portable Document Format) documents.
In~\cite{Tong2017}, Tong, \emph{et al.}, study the effects of iteratively altering malicious PDFs to avoid detection.
Hu and Tan~\cite{HuAdversarialMalwareGan}
propose a generative adversarial network (GAN) for crafting adversarial, malicious Android executable files.

Others have investigated defenses against adversarial malware attacks. Grosse, \emph{et al.},~\cite{Grosse2017a} analyze the distillation defense
for a \textit{static} analysis-based, deep malware classification system which only classifies the raw content of the file without execution.
In~\cite{Grosse2017b}, Grosse, \emph{et al.}, propose a statistical test for detecting adversarial malware examples.

In this paper, we implement and study several adversarial learning-based attacks and defenses for \textit{dynamic} analysis-based, deep learning malware classification systems.
All classification models employ deep neural networks (DNNs).
We study six different strategies of crafting adversarial malware samples based on the removal of malicious features and the addition of benign features.
We evaluate three different defenses for these attacks including the distillation defense as well as the ensemble defense. We also propose and analyze a new weight decay defense.
Results show that the ensemble defense outperforms the other two defenses by a significant margin.
Most models yield a similar classification accuracy compared to their baseline systems, which satisfies a key goal of defensive adversarial learning that the defense does not negatively
affect the overall detection capability.
Finally, while adding additional hidden layers to a neural model only improves the accuracy in a few scenarios,
we demonstrate that a deep neural network offers much better resiliency to adversarial samples compared to its shallow baseline model counterpart.
Furthermore, the resiliency continues to increase as the number of hidden layers in the DNN increases.
A summary of the main contributions of this work includes:
\begin{itemize}
\item{ We are the first to study the efficacy of the distillation defense for dynamic analysis-based, deep malware classification.}
\item{ We propose the weight decay defense and analyze its performance in the context of  malware classification.}
\item{ We show that the ensemble defense is superior in the context of deep malware classification.}
\item{ We demonstrate that adding additional hidden layers significantly increases the resiliency to adversarial attacks.}
\end{itemize}

\section{System Overview and Threat Model}
\label{sec_threat}
\begin{figure*}[!tbh]
\centering
%\vspace{-0.4in}
{\label{}\includegraphics[trim = 0.5in 1.5in 0.5in 2.5in,clip,width=1.75\columnwidth]{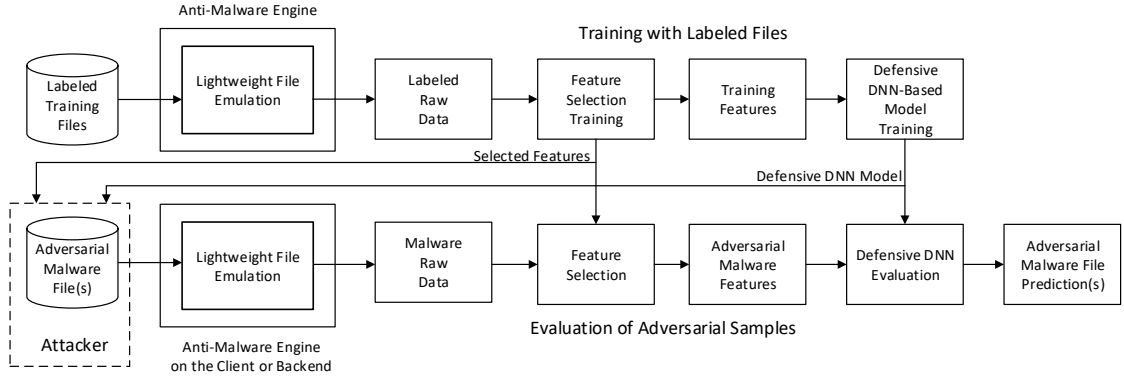}}
\caption{Overview of the adversarial attack and defense of a dynamic analysis-based malware classification system.}
\label{fig:system}
\end{figure*}
In this section, we provide a high-level overview of the defender's training and evaluation systems as well as the threat model which includes the
assumptions about the attacker and the detection strategies.

%\subsection{System Overview}
\textbf{System Overview:}
The system overview is depicted in Figure~\ref{fig:system}.
The original data for this study was collected by scanning a large collection of Windows portable executable (PE) files with a production version
of a commercial anti-malware engine which had been modified to generate two sets of logs for each file including unpacked file strings and system
API (application protocol interface) calls including their parameters. Before an unknown file is executed on the actual operating system,
the anti-malware engine fist analyzes
the file with its lightweight emulator which induces the dynamic behavior of the file. The first log file that is generated
during emulation is a set of unpacked file strings. Typically, a malware file is packed, or encrypted, to make it difficult to
reverse engineer by malware analysts.  During emulation, text strings, which are included in the PE files, are unpacked and
written to the system memory. The emulator's system memory is next scanned to recover null terminated objects which include the original text strings.
In addition, the engine also logs the sequence of API calls and their parameters which are generated during execution.
This sequence provides an indication of the dynamic behavior of the unknown file.

From these two log files, we generate three sets of sparse binary features for our deep learning models. We consider each distinct, unpacked file string as
a potential feature. Two sets of features are derived from the system call data. First, we generate a potential feature for each distinct value of an API call
and input parameter value for a specific input position. Second, we generate all possible combinations of API trigrams (\emph{i.e.}, (k) API call, (k+1) API call, (k+2) API call)
as a feature which represents the local behavior of the file.

There are tens of millions of potential features which  are generated from the three sets of raw features. Since the neural network cannot process this extremely large set of data,
we utilize feature selection using mutual information~\cite{Manning09} in order to reduce the final feature set to 50,000 features. If any of these final features are
generated during emulation, the corresponding feature will be set to 1 in the sparse, binary input feature for that file.
This set of feature vectors is then used to train the deep learning model which has been enhanced to defend against adversarial attacks.

We assume the attacker has knowledge of the selected features and the trained DNN model. With this information, they are able to craft adversarial malware
samples which are processed by the anti-malware engine and the identical inference engine. The goal of the attacker is for their malware sample to have
a benign prediction.

%\subsection{Threat Model}
\textbf{Threat Model:}
We follow earlier work~\cite{papernot2015distillation} and assume that the attacker has access to all of the model parameters and operating thresholds. For
an ensemble classification system, we assume that the attacker has obtained all parameters and threshold values for each classifier in
the ensemble. This is the most challenging scenario to protect.
Once the attacker has successfully obtained of the parameters for the model or ensemble of models, we assume they implement
the Jacobian-based strategies proposed in~\cite{papernot2015distillation,papernot2015limitations} to determine the ranking of important malicious and benign features.

Modern anti-malware systems consist of two main components: an anti-malware client on the user's computer and a backend web service which processes queries
from all of the individual anti-malware clients.
It would be difficult and most likely require a successful spearphishing campaign
to obtain any classification models running in a backend web service. However it would be much easier to reverse engineer a malware classifier's
parameters and threshold values running on a user's client computer.

All of the data is generated by the anti-malware engine's emulator running in a virtual machine without external network access. We assume that
malware does not detect that it is being emulated and halt all malicious activity. We further assume that the malware does not alter its behavior due to the lack of external
internet access.

Finally, in several of the attack strategies proposed in the next section, we assume that the malware author can remove key features related to malicious
activity (\emph{i.e.}, malware features) while maintaining its ability to achieve the desired malicious objective.
Since most malware is either packed or encrypted, our analysis is based on the \textit{behavior} of the malicious code, and we use a dataset of over 2.3 million malware and benign
files in this study, it is impossible for us to actually modify the malware to remove malicious features.
Removing important malicious content may actually transforms the malware into a benign file.
However, attackers often employ metamorphic
strategies to use alternate code paths to reach the desired malicious objective~\cite{Crandall05}.
In order to continue to perform its desired malicious behavior, we assume the attacker has the ability to engineer an alternative attack strategy. For example,
instead of writing a value to the registry, the attacker may choose to instead write important
data to a local file or memory. In other cases, the attacker may re-implement key functions of the operating system. We, therefore, assume the attacker
has the ability to effectively remove malicious features by re-implementing the key pieces of the malware's code
related to the most important malicious features.

\section{Baseline DNN Malware Classifier}
\label{sec:baseline}
\begin{figure}[!tbh]
\centering
%\vspace{-0.4in}
{\label{}\includegraphics[trim = 6.0in 1.0in 2.0in 2.0in,clip,width=0.6\columnwidth]{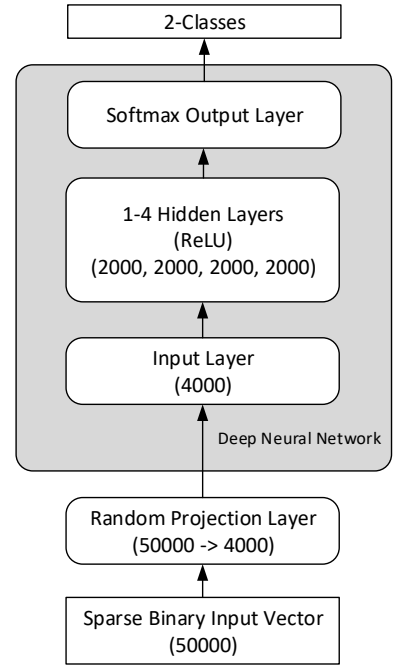}}
\caption{Model of the baseline deep neural network malware classifier.}
\label{fig:baseline}
\end{figure}

Before discussing the strategies for crafting and defending against adversarial samples, we first review the baseline deep neural network malware classifier which
is illustrated in Figure~\ref{fig:baseline}.
We follow earlier work in~\cite{dahl2013large} and use a sparse random projection matrix~\cite{li2006} to reduce the input feature dimension from 50,000 to 4,000 for the DNN's input layer.
The sparse
random projection matrix $R$ is
initialized with 1 and -1 as
\begin{equation}
Pr(R_{i,j}=1)=Pr(R_{i,j}=-1)=\frac{1}{2\sqrt{d}}
\end{equation}
where $d$ is the size of the original input feature vector.
All hidden layers have a dimension of 2000. Following~\cite{Huang2016}, we use the rectified linear unit (ReLU) as the activation
function, and dropout~\cite{Srivastava:2014:DSW:2627435.2670313} is utilized with the dropout rate set to 25\%.
All inputs to the DNN are normalized to have zero mean and unit variance.
The output layer employs the softmax function to generate probabilities for the output predictions:
\begin{equation}
softmax(x) = exp(x)/ \sum\limits_{i=1}^c (exp(x_i)).
\label{eq:softmax}
\end{equation}

\section{Crafting Adversarial Samples}
\label{sec_craft}
In this section, we describe six iterative strategies for crafting adversarial samples.
Essentially, the attacker's strategy is to first discover features that have the most influence on the classification output, and then alter their malware to control these
features.
The Jacobian, which is the forward derivative of the output with respect to the original input, has been
proposed in~\cite{papernot2015distillation,papernot2015limitations} as a good criterion to help determine these features.
For a malware classifier, the prediction
output indicates that an unknown file is either malicious or benign.
Thus, the attacker's goal is to alter (\emph{i.e.}, perturb) the important features such that the malware classification model incorrectly predicts that a malicious file is benign.
To compromise the malware classifier, the attacker can modify their malware to decrease the number of features that are important for a malware prediction, increase the number of features that lead to a benign prediction,
or both.

For each iterative attack strategy that simulates an attacker modifying their malware, we alter one feature
during each iteration and then re-evaluate the Jacobian with respect to the perturbed sample.
We analyze six strategies to craft adversarial samples. The first three methods use the Jacobian information~\cite{papernot2015distillation,papernot2015limitations}  to identify which features to alter:

(1) dec\_pos, \emph{i.e.}, disabling the features that would lead the classifier to predict that an unknown file is malware based on the Jacobian of the classification output with respect to the original input features. We define a feature to be a positive feature if the
Jacobian with respect to the feature is positive. We call these features \emph{positive features} since they are the key indicators of malware behavior.

(2) inc\_neg, \emph{i.e.}, enabling the features that would lead a classifier to predict that an unknown file is benign. These features are called \emph{negative features} with respect to the malware class. A negative feature has a positive
Jacobian with respect to the benign class.

(3) dec\_pos + inc\_neg, \emph{i.e.}, alternatively disabling one positive feature for one iteration and then enabling one negative feature in the next iteration. This strategy investigates whether there is any synergy between removing malicious content and adding benign features in a round robin fashion.

In contrast to the above methods that use the Jacobian information, we also include three, similar ``randomized'' strategies that do not use the Jacobian for comparison. For these additional algorithms, we randomly select positive features to disable or negative features to
enable instead of selecting them using the rank of the Jacobian's forward derivatives. Thus, the additional strategies include:
(4) randomized dec\_pos,
(5) randomized inc\_neg\_random, and
(6) randomized dec\_pos + inc\_neg.

\section{Knowledge Distillation}
In this section, we review the basics of knowledge distillation, which is used in the next section in one of the defensive mechanisms.
Knowledge distillation is the procedure to distill the knowledge learned in one model (teacher model) into another model (student model) which is usually
smaller in size. The student model, which mimics the performance of teacher model, can then be deployed to save computational cost.
Dark Knowledge \cite{hinton2015distilling} is proposed by Hinton, \emph{et al.}, to improve the model's distillation performance. Instead
of using the predicted hard labels as training targets, dark knowledge uses the predicted probability $p=[p_1, p_2, ..., p_c]$ ($\forall i,  0 < p_i < 1$, and $p_i$ is
the probability of a sample being predicted to belong to class $i$) of the teacher model as the training target. The probability scores
are also known as soft targets, in contrast to the hard targets of the original  \{0, 1\} labels. Dark knowledge argues that the
probability scores capture the relationship between classes.
Taking the MNIST dataset for example, digits 1 and 7 appear more
similar than 1 and 8, so the prediction of a hand-written digit with a ground truth of 1 is usually classified with a higher
probability of 7 than 8.
By utilizing the relationship between
classes, the teacher model communicates more information to the student model, which helps train the student model to better
mimic the complex non-linear function learned by the teacher model.

An issue with using the standard softmax function in (\ref{eq:softmax}) in the final output layer of the baseline DNN is that it causes the output
probability scores to concentrate on one class, thereby reducing the correlation between the output classes. To increase the output class correlation,
Hinton, \emph{et al.},~\cite{hinton2015distilling} propose using a temperature parameter, $T$, to normalize the logit value. Higher values of $T$ cause the probability
scores to be distributed more evenly thereby better reflecting the correlation between classes.
Specifically,
\begin{equation}
p = softmax(z/T) \\
\end{equation}
where $p\in \mathbb{R}^{c \times 1}$ are the output probability scores \emph{w.r.t.} each class, softmax (\ref{eq:softmax}) is a function
which is usually applied to a vector to generate probability scores, $z \in \mathbb{R}^{c \times 1}$ are the logit values (\emph{i.e.}, the output
of the previous layer which is input to the softmax function) output by a deep neural network $f(x)$ applied on an input $x$.  Intuitively when $T$ is large, the
difference between the maximum and minimum normalized logit values is small. Thus, the output
probability values will be pushed towards a more uniform distribution.

If the student model matches the soft targets on a large transfer dataset, then we can say that the student model \textit{distills}
most of the knowledge stored in the  larger teacher model. Note that the transfer set does not need to be
constrained to the original data used for training, but could be any data. Therefore, we can formulate the model distillation problem as soft
target alignment via the cross-entropy loss between the probability scores of the student model and teacher model.

In the distillation process, the student model typically uses the same temperature as the teacher model. During training, the temperature needs to be tuned for the best performance.
When deploying the student model,
the standard softmax function (\ref{eq:softmax}) should be utilized to set the probability scores back to their normal values.

\section{Defensive Methods}
\label{sec_method}
In this section, we review three methods for defending against adversarial attacks including the distillation, ensemble, and weight decay defenses.
Although the distillation and ensemble defenses have been previously proposed, the weight decay defense is new.
Only the distillation defense has been previously explored to defend against adversarial attacks in malware detection applications, and this work was done in the context of static malware classification~\cite{Grosse2017a}.

%\subsection{Distillation Defense}
\textbf{Distillation Defense:}
The first defense we study is the distillation defense~ \cite{papernot2015distillation,Grosse2017a} where the model model is trained using knowledge distillation.
As discussed previously, knowledge distillation is typically used to distall the knowledge learned from a large model into a smaller network making the smaller model more efficient
in terms of its memory, energy, or processing time in deployment. However, in adversarial learning, the goal is to make the distilled model more robust to adversarial perturbations, instead of focusing on compressing the network size.
%The method has been used in the vision scenario as in \cite{papernot2015distillation}. In this work, we analyze the effect of distillation as a defense in the malware detection space.

The motivation of using model distillation as a defense mechanism is that with a higher temperature during the distillation process, the error surface of the learned model can be smoothed. We denote the function learned by the neural network model as $F$. During the inference stage, the feature vector is input into the trained network and transformed into logit scores $z \in \mathbb{R}^{c \times 1}$. Then a
softmax function is used to convert those scores into probabilities with respect to each class. Mathematically, the Jacobian's forward derivative of the output
with respect to the input can be calculated as follows~\cite{papernot2015distillation,papernot2015limitations}.
For notational clarity, we denote the denominator of the softmax function as $h(x) = \sum\limits_{k=1}^c (exp(z_k)/T)$, where $T$ is the temperature used during distillation.
Thus, we have:
\begin{eqnarray}
& \frac{\partial F_i}{\partial x_j}  &= \frac{\partial}{\partial x_j} (\frac{e^{z_i/T}}{h(x)}) \nonumber \\
&& = \frac{1}{h^2(x)} (\frac{\partial e^{z_i(x)/T}}{\partial x_j} h(x) - e^{z_i/T} \frac{\partial h(x)}{\partial x_j}) \nonumber \\
&& = \frac{1}{T}\frac{e^{z_i/T}}{h^2(x)} (\sum\limits_{k=1}^c (\frac{\partial z_i}{\partial x_j} - \frac{\partial z_k}{\partial x_j})e^{z_k/T}).
\label{eqn:dist1}
\end{eqnarray}
From (\ref{eqn:dist1}), we see that as the derivative becomes smaller with higher temperature, the model is less sensitive to adversarial perturbations.

%\subsection{Ensemble Defense}
\textbf{Ensemble Defense:}
The ensemble defense for extraction attacks and evasion attacks has been recently proposed by several authors~\cite{Kantchelian2016,Tramer2016, Tramer2017}
for tree ensemble classifiers. In this work, we study the ensemble defense with neural networks.
The idea behind the ensemble
defense is intuitive.  It may be easy for an attacker to craft adversarial samples to compromise an individual detection model, but it is much more
difficult for them to create samples which fool a set of models in an ensemble with different properties.
We employ a ``majority vote'' ensemble defense
in this work. We first train an ensemble with $E$ classifiers where $E$ is an odd number.
During prediction, an unknown file is predicted to be malware if
the majority (\emph{i.e.}, $> E/2$) of the classifiers predict that the file is malicious.

%\subsection{Weight Decay Defense}
\textbf{Weight Decay Defense:}
The third defense we propose and study is the weight decay defense.
Weight decay is typically used to prevent overfitting of machine learning models. The $\ell_2$ norm of a weight matrix is defined as the square sum of all the elements. By adding an $\ell_2$ penalty of the model weights in the objective function during optimization, the model is encouraged to prefer smaller magnitude weights since large values are penalized by the objective function.

With a smaller magnitude of weights, the function parameterized by the neural network is smoother, and therefore, changes in the input space lead to smaller changes in the output of a deep learning model. We conjecture that weight decay could help alleviate the vulnerability of a deep learning system against adversarial attacks.

\section{Experimental Results}
In this section, we evaluate the adversarial defenses against the different attack strategies described in the previous sections.
We first describe some details related to data preparation and experimental setup.
We then present
the performance of the baseline classification system which does not employ any defenses. Finally, we evaluate the results for the distillation, weight decay, and the ensemble defenses.

%\subsection{Data Preparation and Setup}
\textbf{Data Preparation and Setup:}
In some cases, multiple files can share the same input vector. Therefore, we only include the first instance of a unique input vector and discard any remaining duplicates.
After de-duplication, we have input data and labels from 2,373,671 files. A file is assigned the label of 1 if it is malware and 0 if it is benign.
We then randomly split the original dataset into a training set, validation set, and test set including 1,523,978, 268,937, and 580,756 files, respectively.

In our training, we implement all models using the
Microsoft Cognitive Toolkit (CNTK)~\cite{CNTK}. All models are derived from the baseline model described Section~\ref{sec:baseline}.
We use the adam optimizer for training where the initial step size is set to 0.1. Training proceeds for each step size
until no further improvement is observed in the validation error. At that point, CNTK halves the step size for subsequent epochs. We train for a maximum of 200 epochs, but
CNTK implements early stopping when no additional improvement in the validation error is observed for a minimum step size of 1e-4.

\begin{figure}[!tbh]
\centering
%\vspace{-0.4in}
{\label{}\includegraphics[trim = 1.25in 3.0in 1.5in 3.0in,clip,width=0.9\columnwidth]{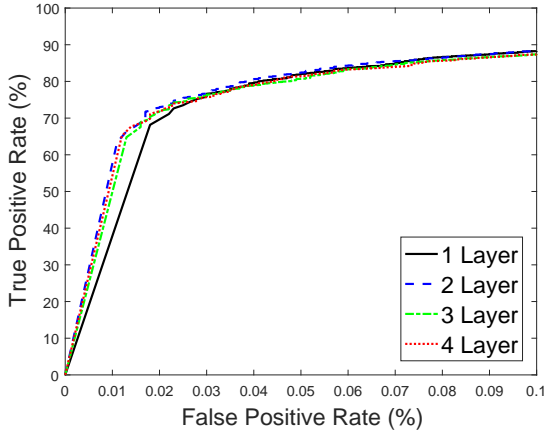}}
\caption{ROC curves of the baseline malware classifier for different numbers of hidden layers.}
\label{fig:Base-ROC}
\end{figure}

%\subsection{Baseline Classifier Perfromance}
\textbf{Baseline Classifier:}
Before investigating the various defenses, we first analyze the performance of the baseline malware classifier useing the receiver operating characteristic (ROC) curves
depicted in Figure~\ref{fig:Base-ROC} for a range of DNN hidden layers, $H$, varying from 1 to 4. Malware classifiers need to operate at very low false positive rates to
avoid false positive detections which may result in the removal of critical operating system and legitimate application files. Thus, our desired operating point is a false positive rate (FPR) of 0.01\%.
While the DNNs with multiple hidden layers offer equivalent performance at higher false positive rates compared to a shallow neural network with one hidden layer, the figures
indicates that the DNNs offer improved performance at very low false positive rates. In particular, the false positive rate of the shallow neural model
immediately jumps to over 0.015\% which is above our desired operating point.

For reference, we next analyze the test error rates of the baseline malware classification system in Table~\ref{table:Base-Test-Error-Deduped-Train-Valid-Test-2}.
As observed
in~\cite{dahl2013large} for a different dataset created for dynamic analysis malware classification, a shallow neural network with a single hidden layer provides the best overall accuracy.
The test error rates in Table~\ref{table:Base-Test-Error-Deduped-Train-Valid-Test-2} are computed with the probability that the file is malicious $p_M \ge 0.5$. This threshold corresponds to operating
points with higher false positive rates on the ROC curves for $H \in {1,2,3,4}$. These higher thresholds explain why the shallow network has a better test error, but the ROC curves indicate
better performance for multiple hidden layers at the same FPR.
%
%
%\begin{comment}
\begin{table}[!hbt]
    \centering
    \begin{small}
    \begin{tabular}{l|c}
    Layers  & Test Error Rate (\%)  \\ \hline
    1       & 1.1378272 \\
    2       & 1.2053255 \\
    3       & 1.1762255 \\
    4       & 1.1619338 \\
    \end{tabular}
    \end{small}
    %\caption{Test error rates for the distillation defense for various temperature settings for the new test set which does not have duplicates in the training, validation or test sets. The test set is the same as in Table~\ref{table:Test-Error-Filtered-Test}}
    \caption{Test error rates of the baseline malware classifier for different numbers of hidden layers.}
    \label{table:Base-Test-Error-Deduped-Train-Valid-Test-2}
\end{table}
%\end{comment}

\begin{figure}[!tbh]
\centering
%\vspace{-0.4in}
{\label{}\includegraphics[trim = 1.25in 3.0in 1.5in 3.0in,clip,width=0.9\columnwidth]{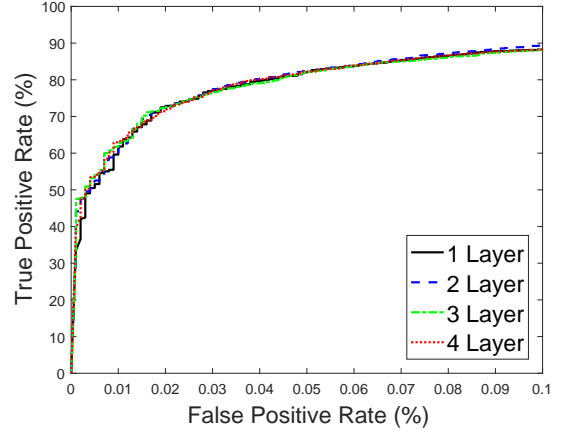}}
\caption{ROC curves of the malware classifiers with the distillation defense with $T=2$ for different numbers of hidden layers.}
\label{fig:T-2-ROC}
\end{figure}

\begin{figure}[!tbh]
\centering
%\vspace{-0.4in}
{\label{}\includegraphics[trim = 1.25in 3.0in 1.5in 3.0in,clip,width=0.9\columnwidth]{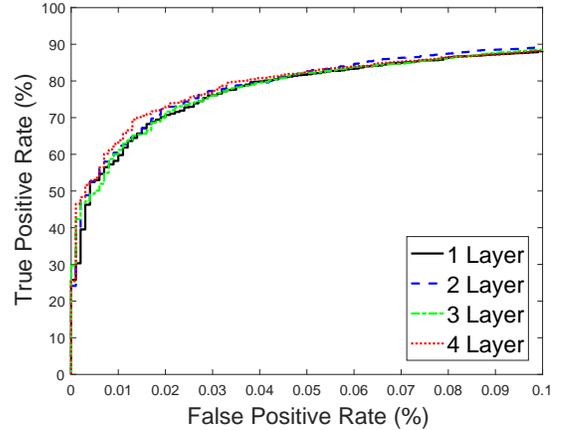}}
\caption{ROC curves of the malware classifiers with the distillation defense with $T=10$ for different numbers of hidden layers.}
\label{fig:T-10-ROC}
\end{figure}

%\subsection{Distillation Defense}
\textbf{Distillation Defense:}
We next analyze the performance of the distillation defense system for all malware and benign files.
The ROC curves of the DNN systems employing the distillation defense are presented in Figure~\ref{fig:T-2-ROC} for temperature setting $T = 2$ and Figure~\ref{fig:T-10-ROC} for $T = 10$.
We make several observations from these figures.
Both systems provide multiple operating points below FPR = 0.01\% which allows better fine-tuning of the models. For the model with $T = 2$, we do not observe any benefit from adding multiple hidden layers.
However, we do get a small lift in the performance for the DNN with 4 hidden layers for $T = 10$. Both systems offer similar performance above an FPR = 0.02\% compared to the baseline classifiers in Figure~\ref{fig:Base-ROC}.


\begin{figure*}[!t]
\centering
\subfloat[$H$ = 1, Baseline]
{\label{}\includegraphics[trim = 1.25in 3.25in 1.25in 3.25in,clip,width=0.70\columnwidth]{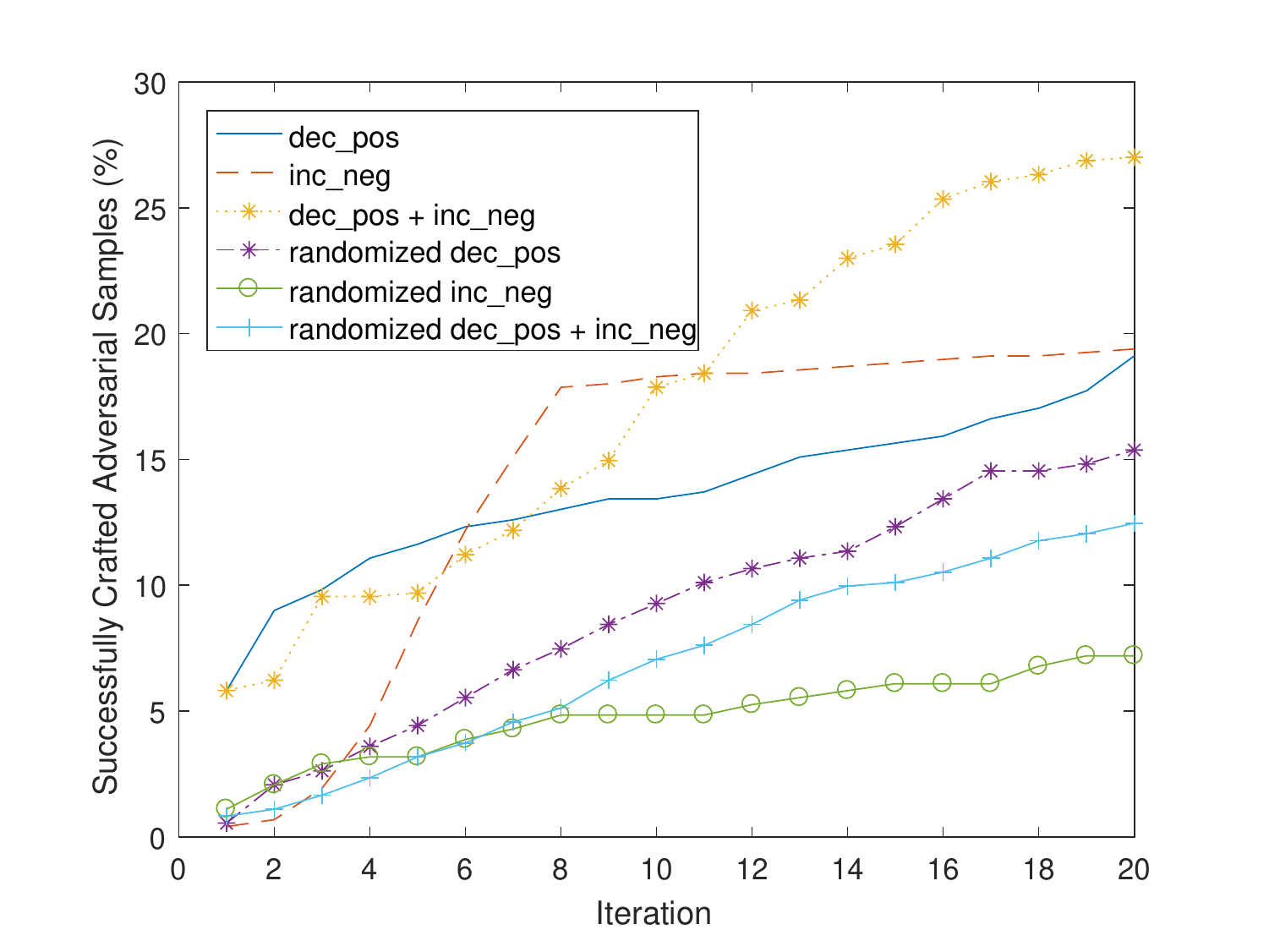}}
%{\label{}\includegraphics[width=0.33\textwidth]{eps/JayTestTrainValidTestDistillationDistinctFeatures_Temp_1_0_MeanVar_Layer_1.eps}}
%\hspace{-1.5in}
\subfloat[$H$ = 1, $T$ = 2]
{\label{}\includegraphics[trim = 1.25in 3.25in 1.25in 3.25in,clip,width=0.70\columnwidth]{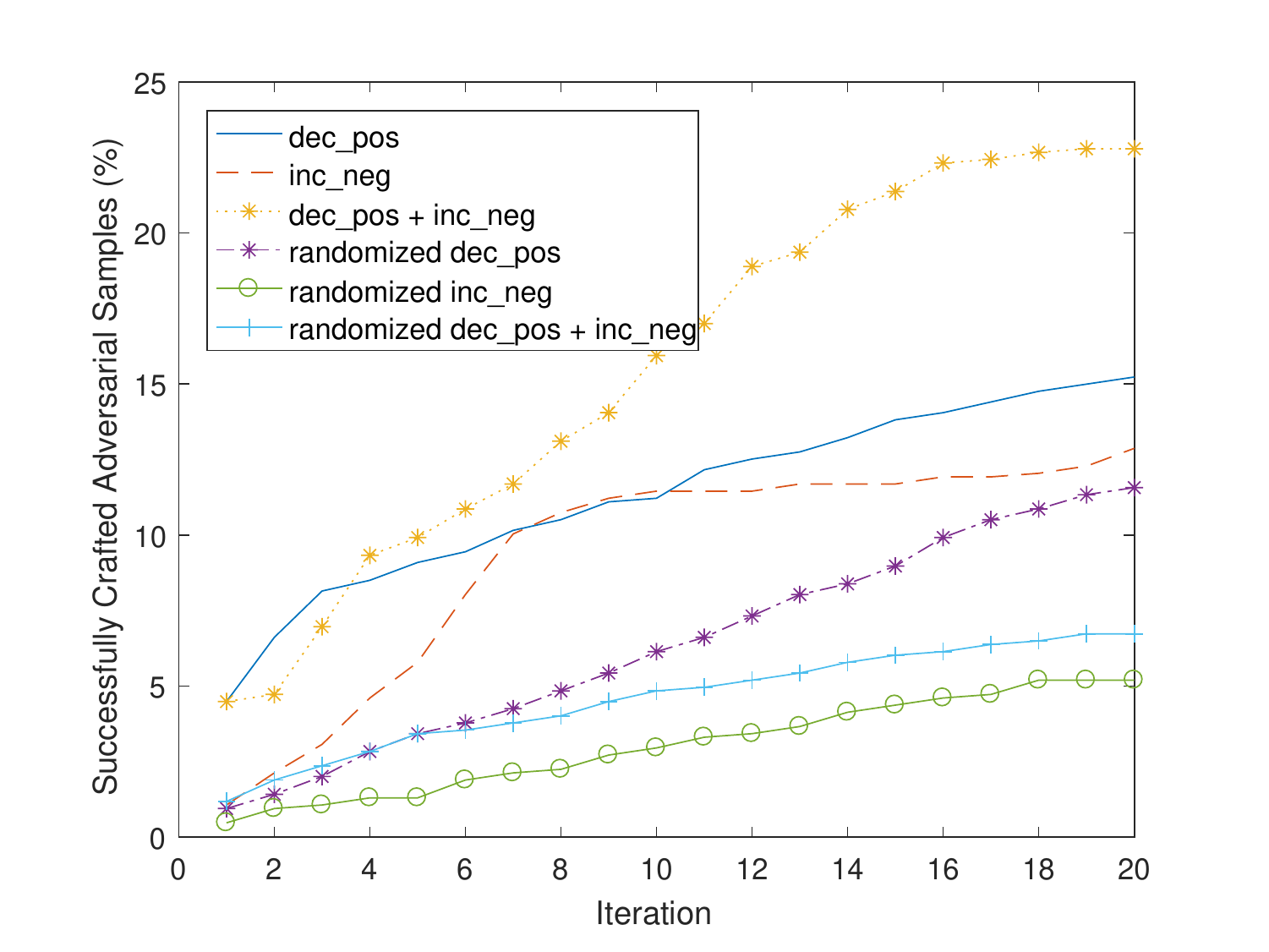}}
%{\label{}\includegraphics[width=0.33\textwidth]{eps/JayTestTrainValidTestDistillationDistinctFeatures_Temp_0_5_MeanVar_Layer_1.eps}}
\subfloat[$H$ = 1, $T$ = 10]
{\label{}\includegraphics[trim = 1.25in 3.25in 1.25in 3.25in,clip,width=0.70\columnwidth]{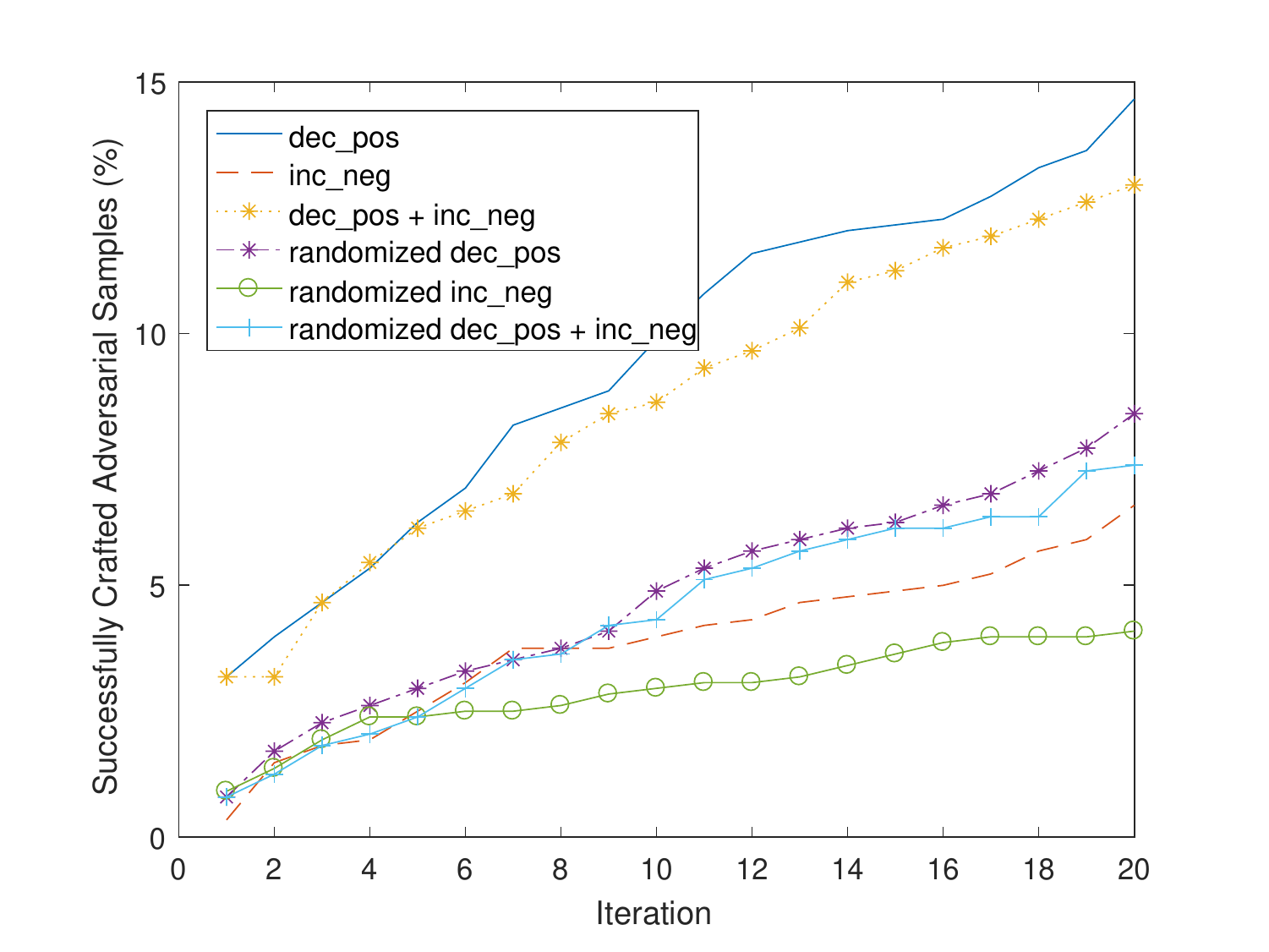}}
%{\label{}\includegraphics[width=0.33\textwidth]{eps/JayTestTrainValidTestDistillationDistinctFeatures_Temp_0_1_MeanVar_Layer_1.eps}}
%\hspace{-1.5in}
\\
\subfloat[$H$ = 2, Baseline]
{\label{}\includegraphics[trim = 1.25in 3.25in 1.25in 3.25in,clip,width=0.70\columnwidth]{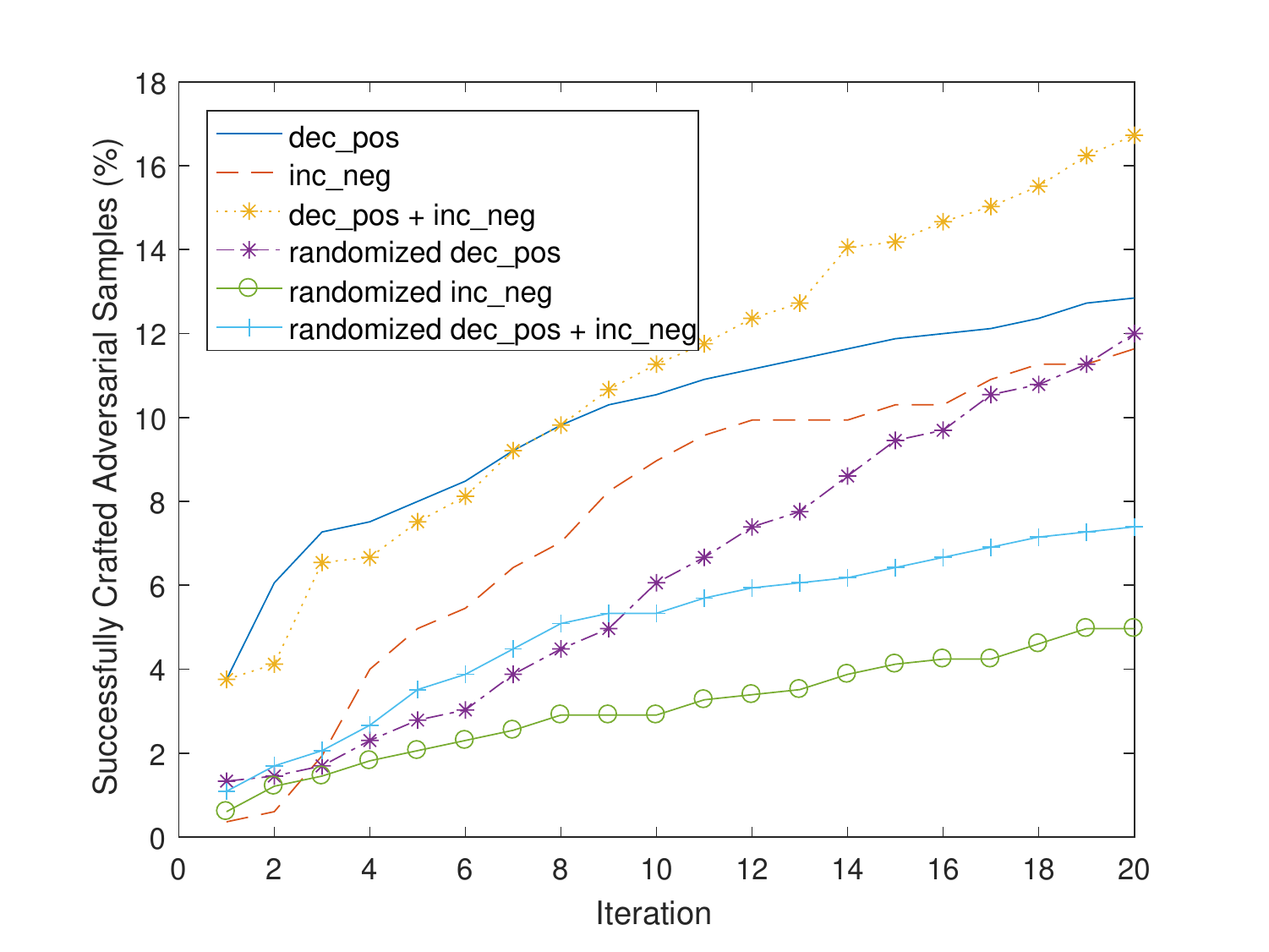}}
%{\label{}\includegraphics[width=0.33\textwidth]{eps/JayTestTrainValidTestDistillationDistinctFeatures_Temp_1_0_MeanVar_Layer_2.eps}}
%\hspace{-1.5in}
\subfloat[$H$ = 2, $T$ = 2]
{\label{}\includegraphics[trim = 1.25in 3.25in 1.25in 3.25in,clip,width=0.70\columnwidth]{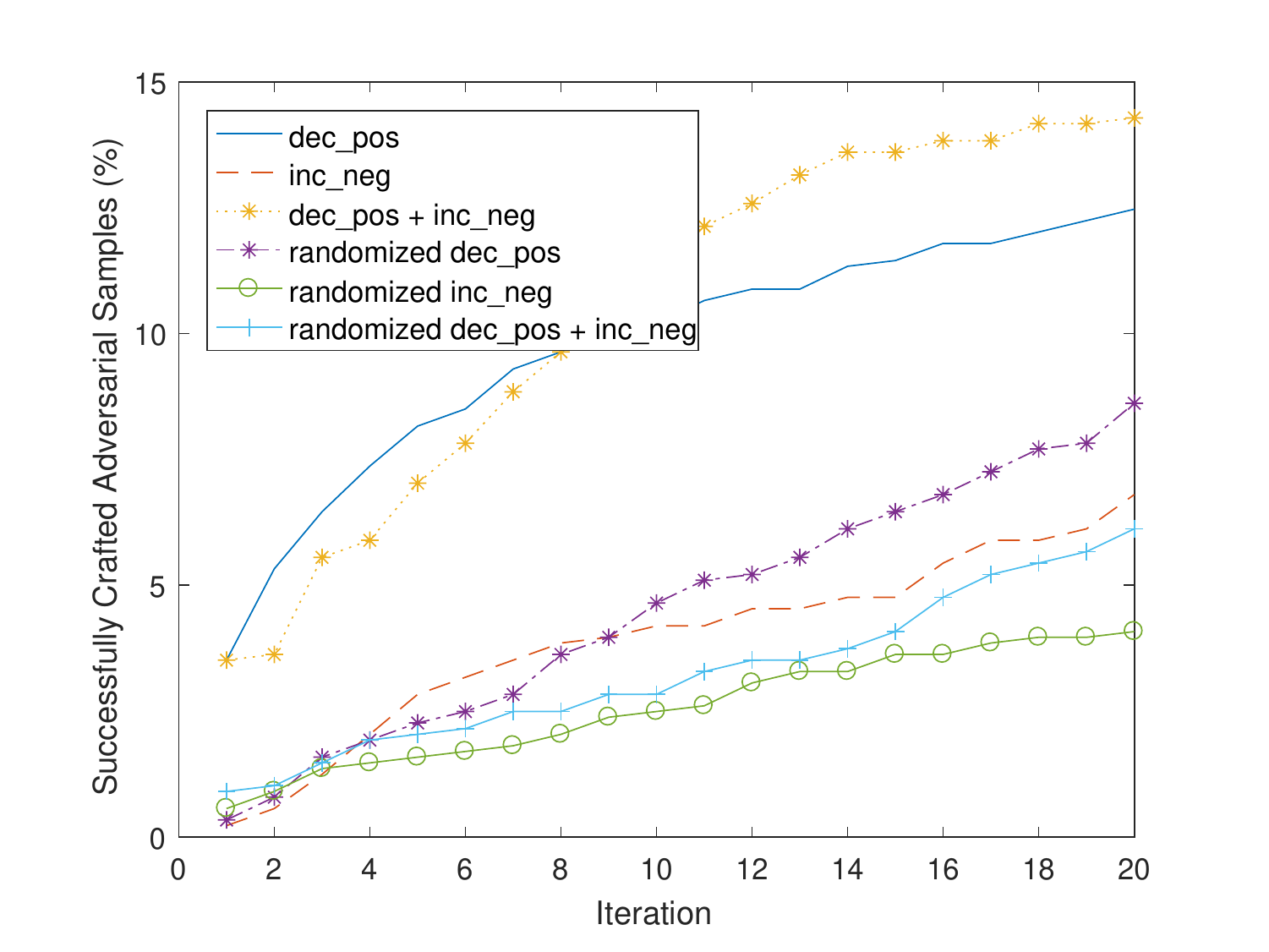}}
%{\label{}\includegraphics[width=0.33\textwidth]{eps/JayTestTrainValidTestDistillationDistinctFeatures_Temp_0_5_MeanVar_Layer_2.eps}}
\subfloat[$H$ = 2, $T$ = 10]
{\label{}\includegraphics[trim = 1.25in 3.25in 1.25in 3.25in,clip,width=0.70\columnwidth]{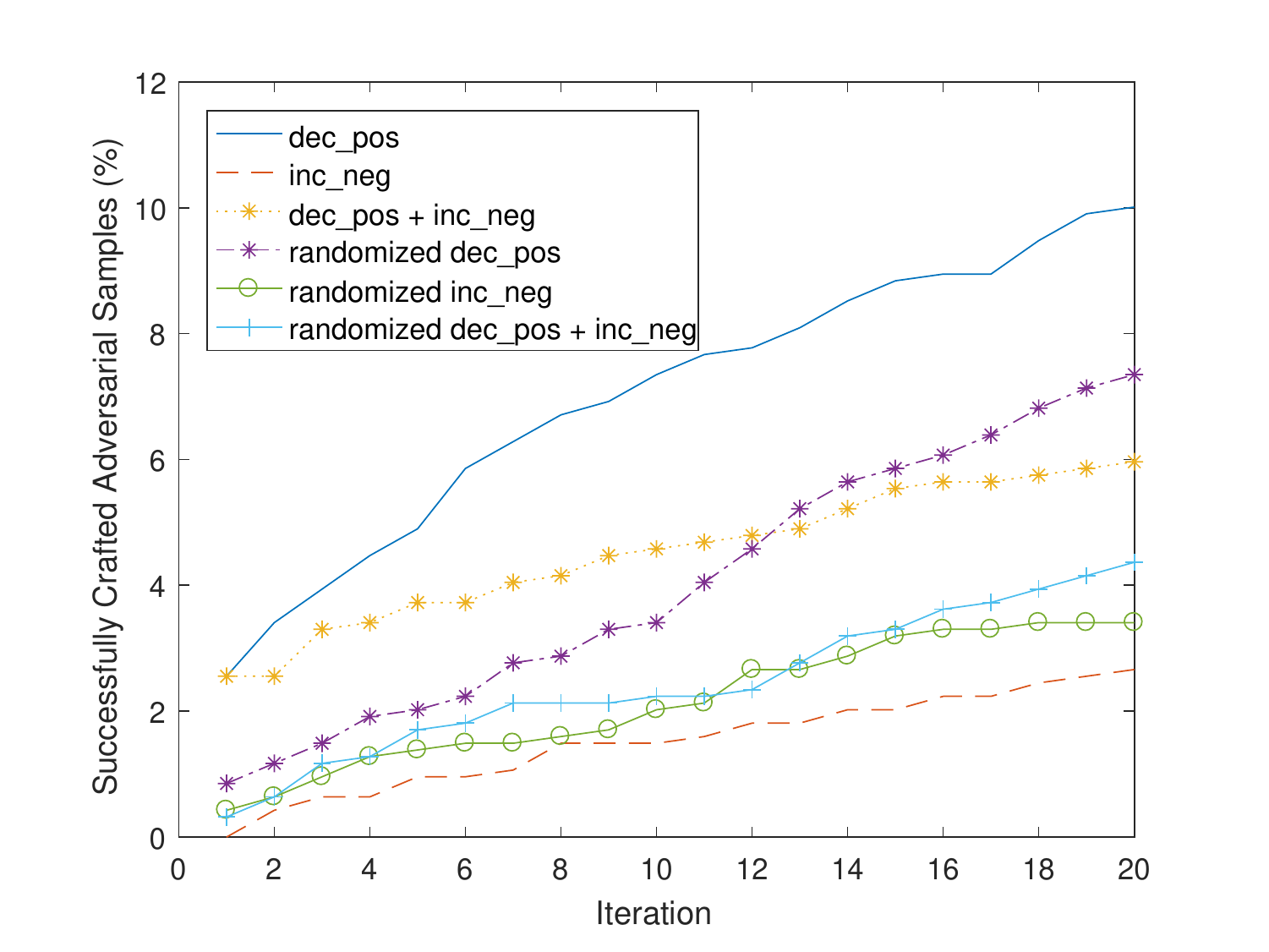}}
%{\label{}\includegraphics[width=0.33\textwidth]{eps/JayTestTrainValidTestDistillationDistinctFeatures_Temp_0_1_MeanVar_Layer_2.eps}}
\\
\subfloat[$H$ = 3, Baseline]
{\label{}\includegraphics[trim = 1.25in 3.25in 1.25in 3.25in,clip,width=0.70\columnwidth]{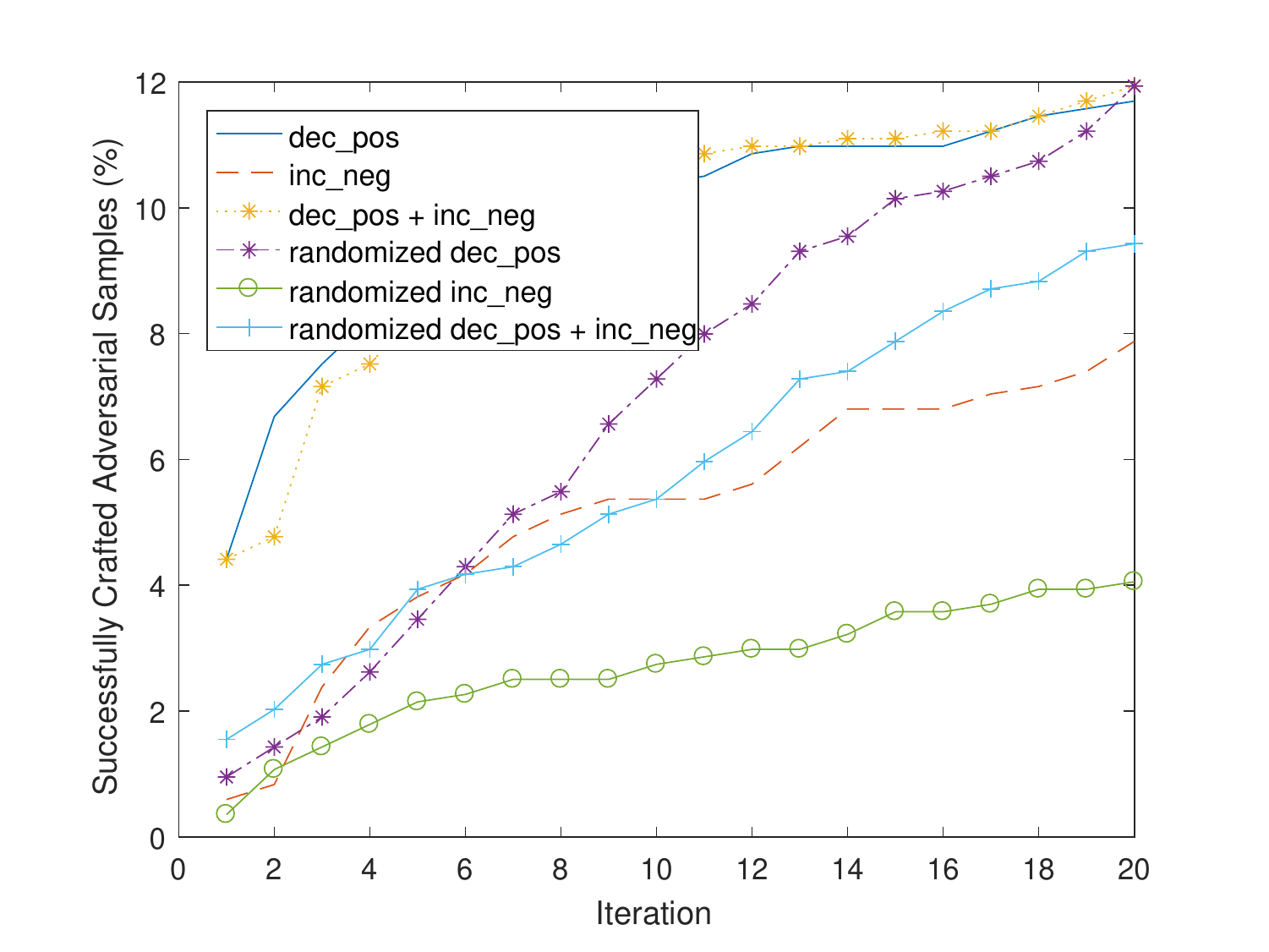}}
%{\label{}\includegraphics[width=0.33\textwidth]{eps/JayTestTrainValidTestDistillationDistinctFeatures_Temp_1_0_MeanVar_Layer_3.eps}}
%\hspace{-1.5in}
\subfloat[$H$ = 3, $T$ = 2]
{\label{}\includegraphics[trim = 1.25in 3.25in 1.25in 3.25in,clip,width=0.70\columnwidth]{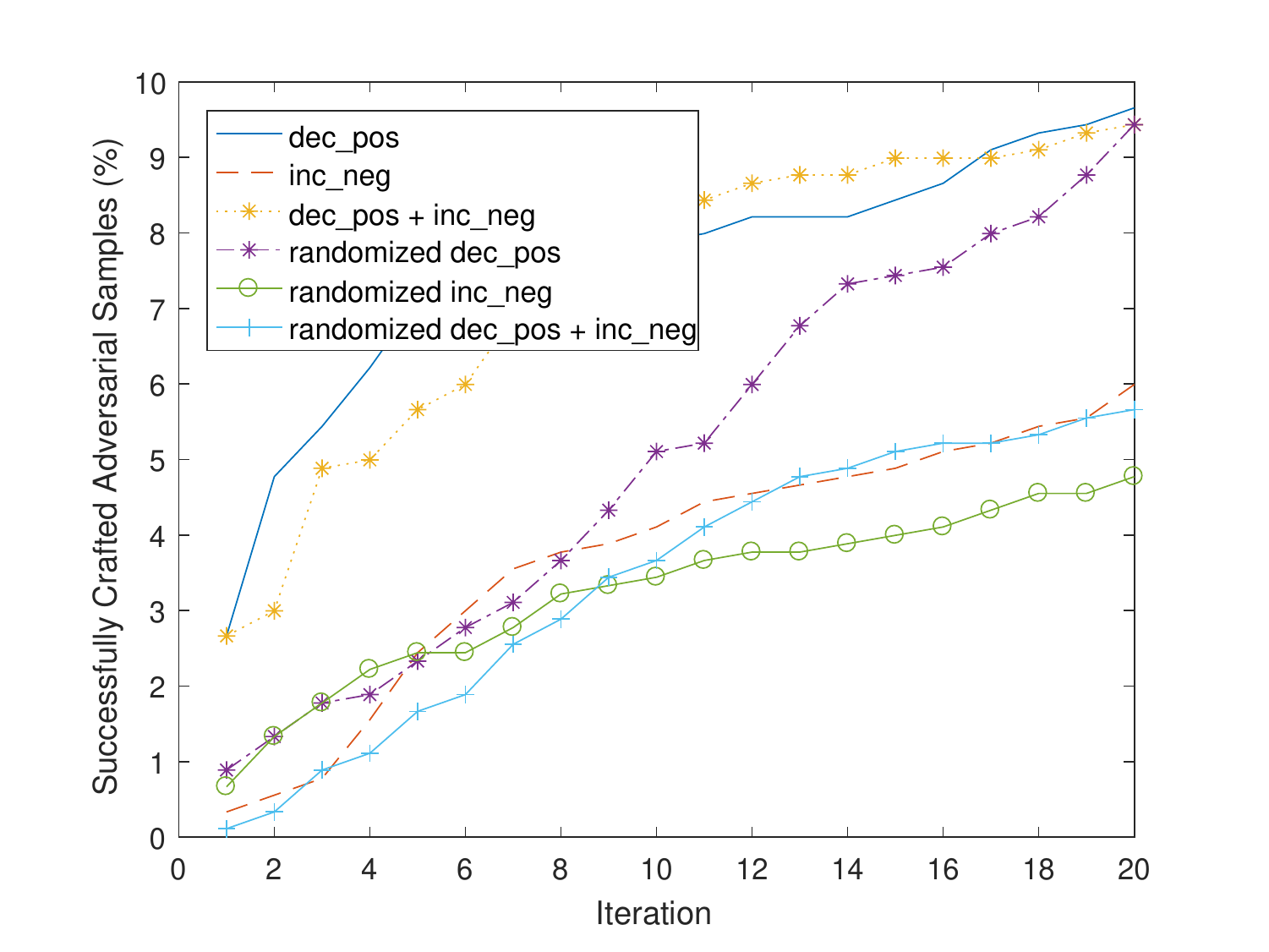}}
%{\label{}\includegraphics[width=0.33\textwidth]{eps/JayTestTrainValidTestDistillationDistinctFeatures_Temp_0_5_MeanVar_Layer_3.eps}}
\subfloat[$H$ = 3, $T$ = 10]
{\label{}\includegraphics[trim = 1.25in 3.25in 1.25in 3.25in,clip,width=0.70\columnwidth]{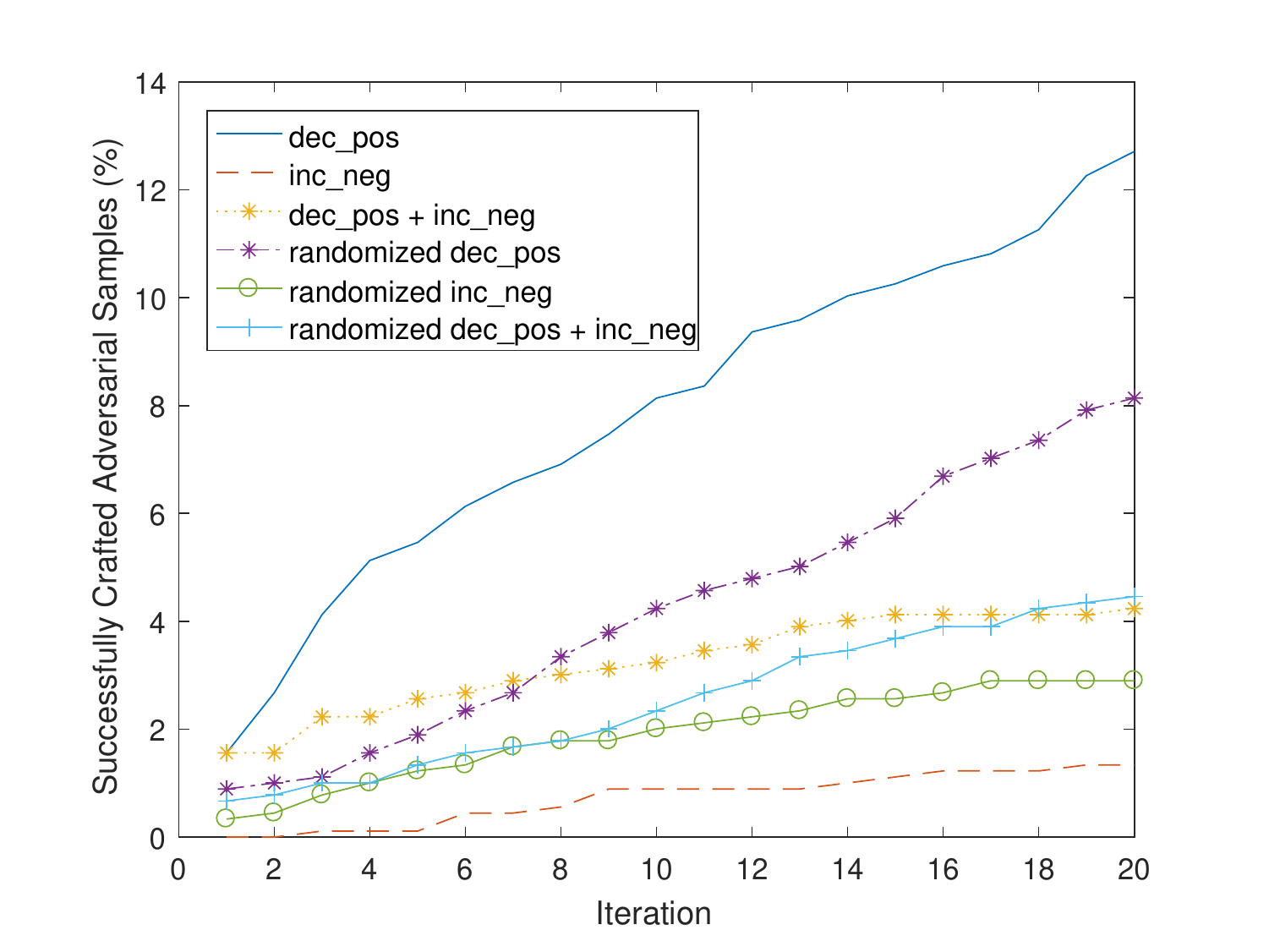}}
%{\label{}\includegraphics[width=0.33\textwidth]{eps/JayTestTrainValidTestDistillationDistinctFeatures_Temp_0_1_MeanVar_Layer_3.eps}}
\\
\subfloat[$H$ = 4, Baseline]
{\label{}\includegraphics[trim = 1.25in 3.25in 1.25in 3.25in,clip,width=0.70\columnwidth]{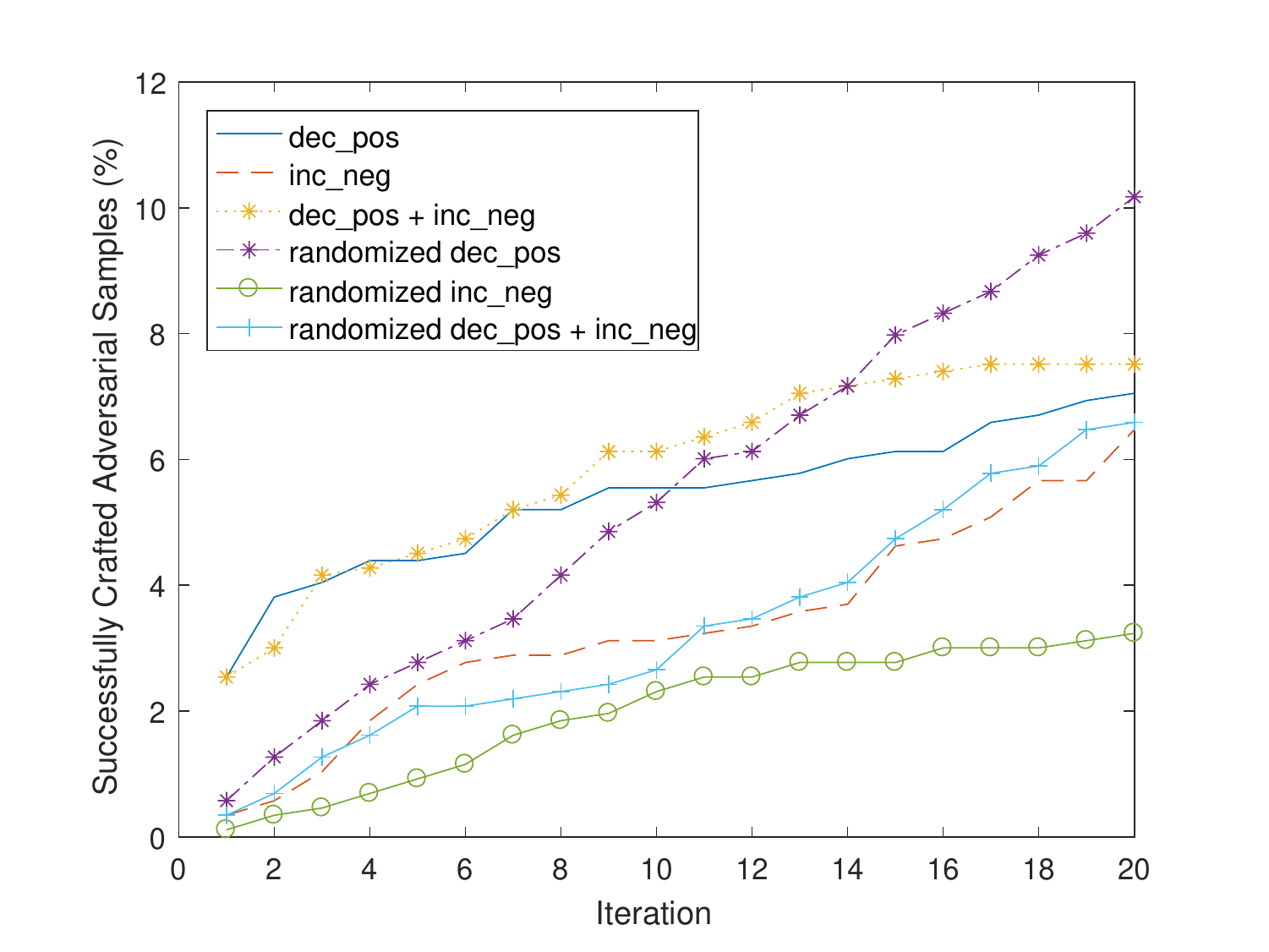}}
%{\label{}\includegraphics[width=0.33\textwidth]{eps/JayTestTrainValidTestDistillationDistinctFeatures_Temp_1_0_MeanVar_Layer_4.eps}}
%\hspace{-1.5in}
\subfloat[$H$ = 4, $T$ = 2]
{\label{}\includegraphics[trim = 1.25in 3.25in 1.25in 3.25in,clip,width=0.70\columnwidth]{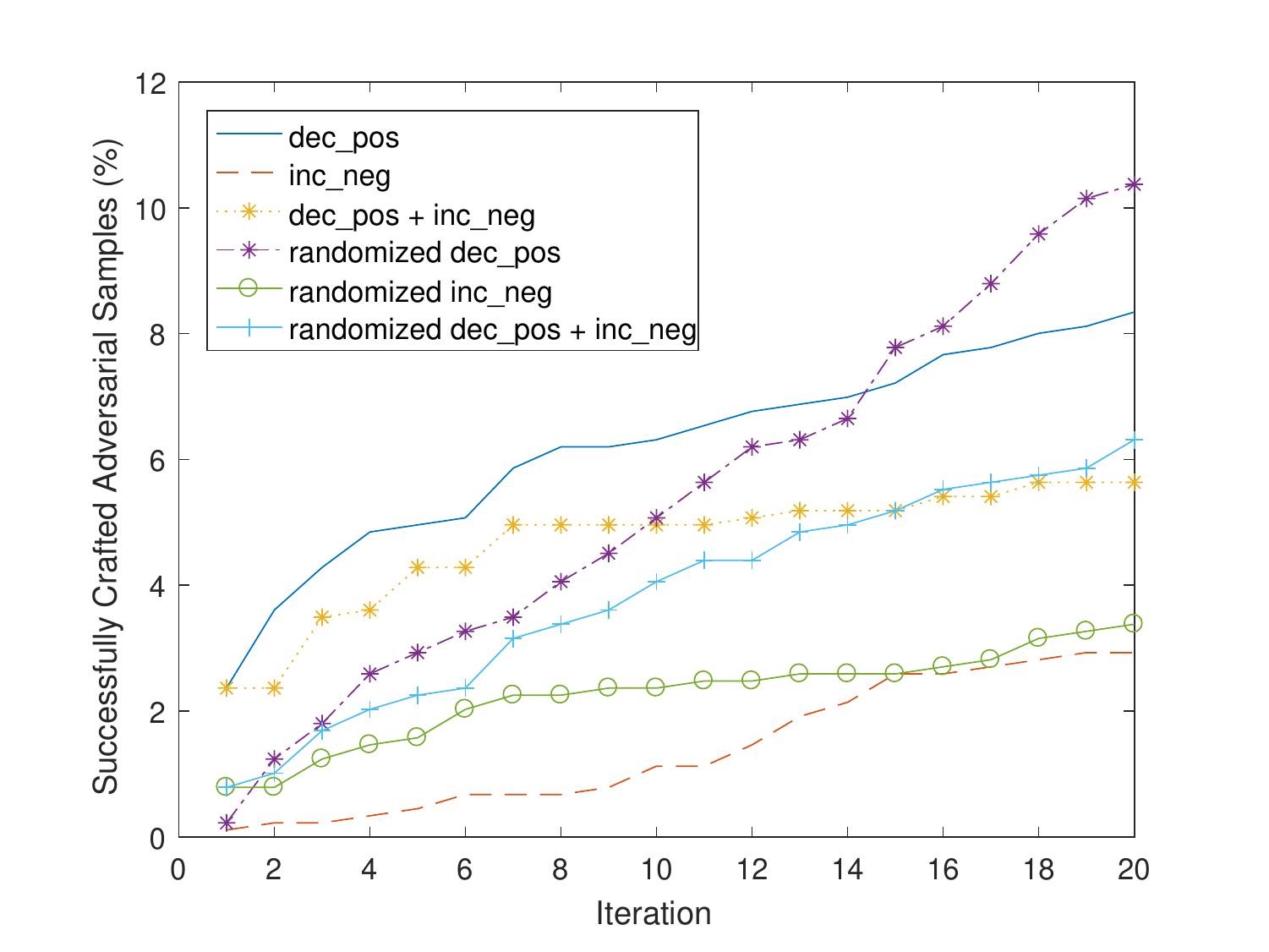}}
%{\label{}\includegraphics[width=0.33\textwidth]{eps/JayTestTrainValidTestDistillationDistinctFeatures_Temp_0_5_MeanVar_Layer_4.eps}}
\subfloat[$H$ = 4, $T$ = 10]
{\label{}\includegraphics[trim = 1.25in 3.25in 1.25in 3.25in,clip,width=0.70\columnwidth]{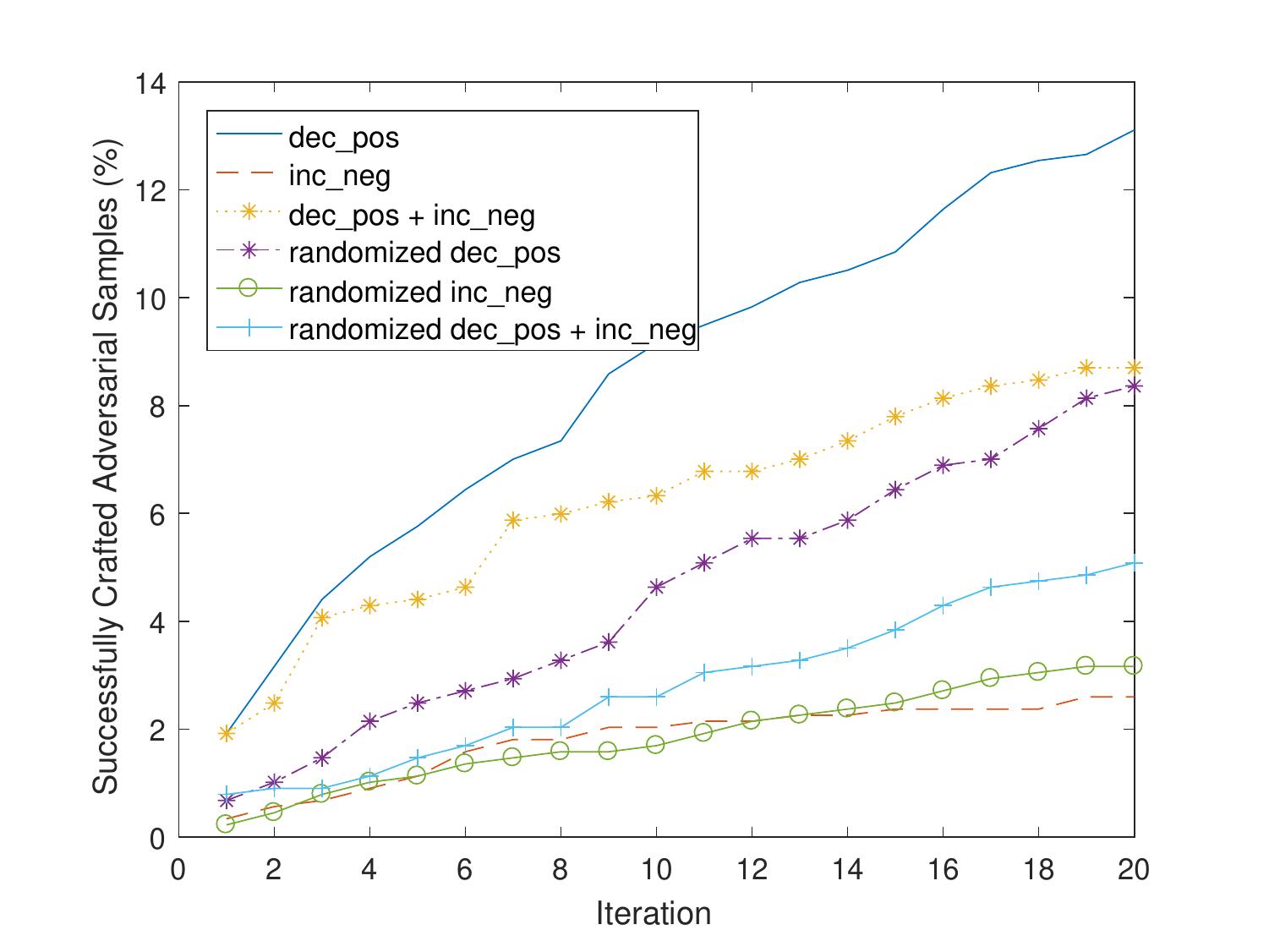}}
%{\label{}\includegraphics[width=0.33\textwidth]{eps/JayTestTrainValidTestDistillationDistinctFeatures_Temp_0_1_MeanVar_Layer_4.eps}}
\caption{Success rates of adversarial samples against the baseline classifer and the using defensive distillation with temperatures, $T \in {2,10}$. Each subfigure shows the results of a DNN with different number of hidden layers, $H$. }
\label{temp}
\end{figure*}

 In Figure~\ref{temp}, we next investigate the effectiveness of the six adversarial sample crafting strategies for the baseline classifiier and distillation defense, with temperatures $T = \in \{2, 10\}$, for model depths $H \in \{1,2,3,4\}$.
In each iteration, a single feature is modified, and the generated sample is evaluated by the trained model to test whether the sample is misclassified.
%Note that we only perturb malware samples such that it can escape from the detection system. The ratio of samples being misclassified by the malware classification model is computed after each iteration.p
From Figure~\ref{temp}, we make several observations.
Generally, the distilled models follow a similar trend with regard to the six strategies for crafting adversarial samples, where dec\_pos and dec\_pos+inc\_neg are the two most effective strategies for the attacker.
With a higher distillation temperature, it becomes much harder to craft adversarial samples for the distilled model. If the same number of features is perturbed, the success rate for crafting adversarial samples is reduced significantly for models distilled with a higher temperature. This result is because the error surface of the distilled model is smoothed for higher temperatures, such that the output is less sensitive with respect to the input.

\begin{figure}[!tbh]
\centering
\hspace{0in}
{\label{}\includegraphics[trim = 1.0in 3.0in 1.0in 3.0in,clip,width=1.0\columnwidth]{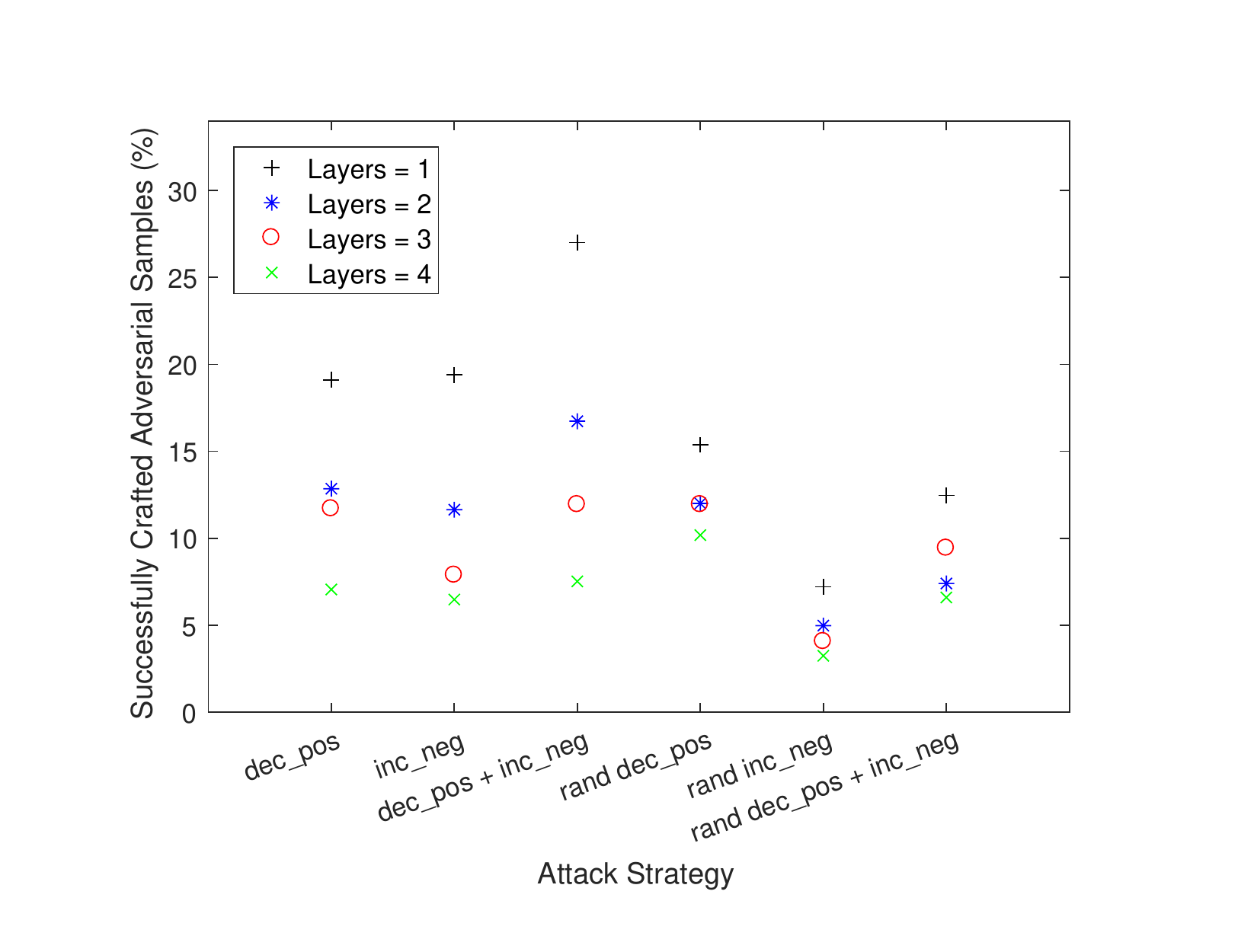}}
\caption{Percentage of successfully crafted adversarial samples for different sample crafting strategies for the baseline model with no defense.}
\label{strategySummaryT1}
\end{figure}

\begin{figure}
\centering
\hspace{0in}
{\label{}\includegraphics[trim = 1.0in 3.0in 1.0in 3.0in,clip,width=1.0\columnwidth]{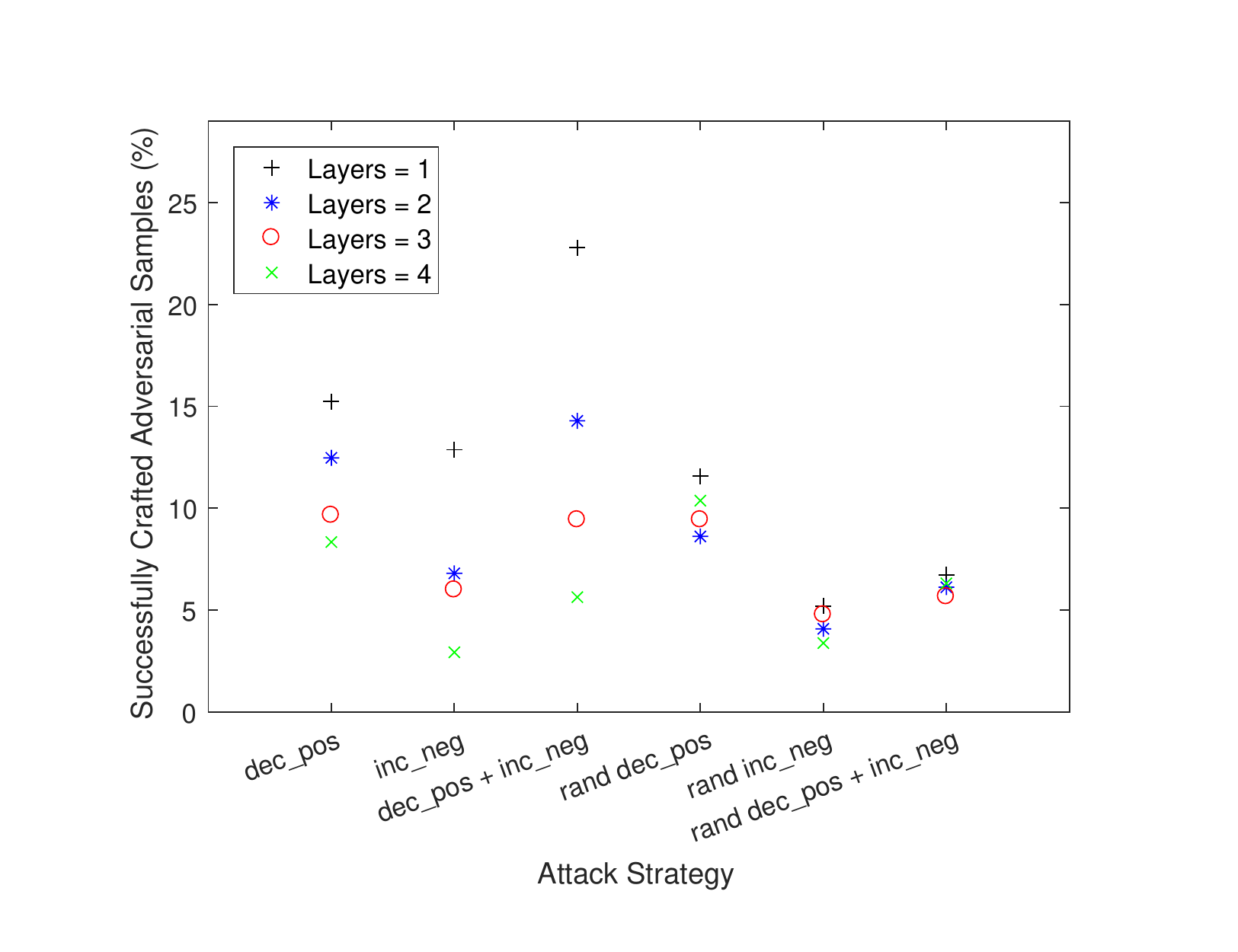}}
\caption{Percentage of successfully crafted adversarial samples for different sample crafting strategies with the distillation defense and $T=2$.}
\label{strategySummaryT2}
\end{figure}

\begin{figure}[!tbh]
\centering
\hspace{0in}
{\label{}\includegraphics[trim = 1.0in 3.0in 1.0in 3.0in,clip,width=1.0\columnwidth]{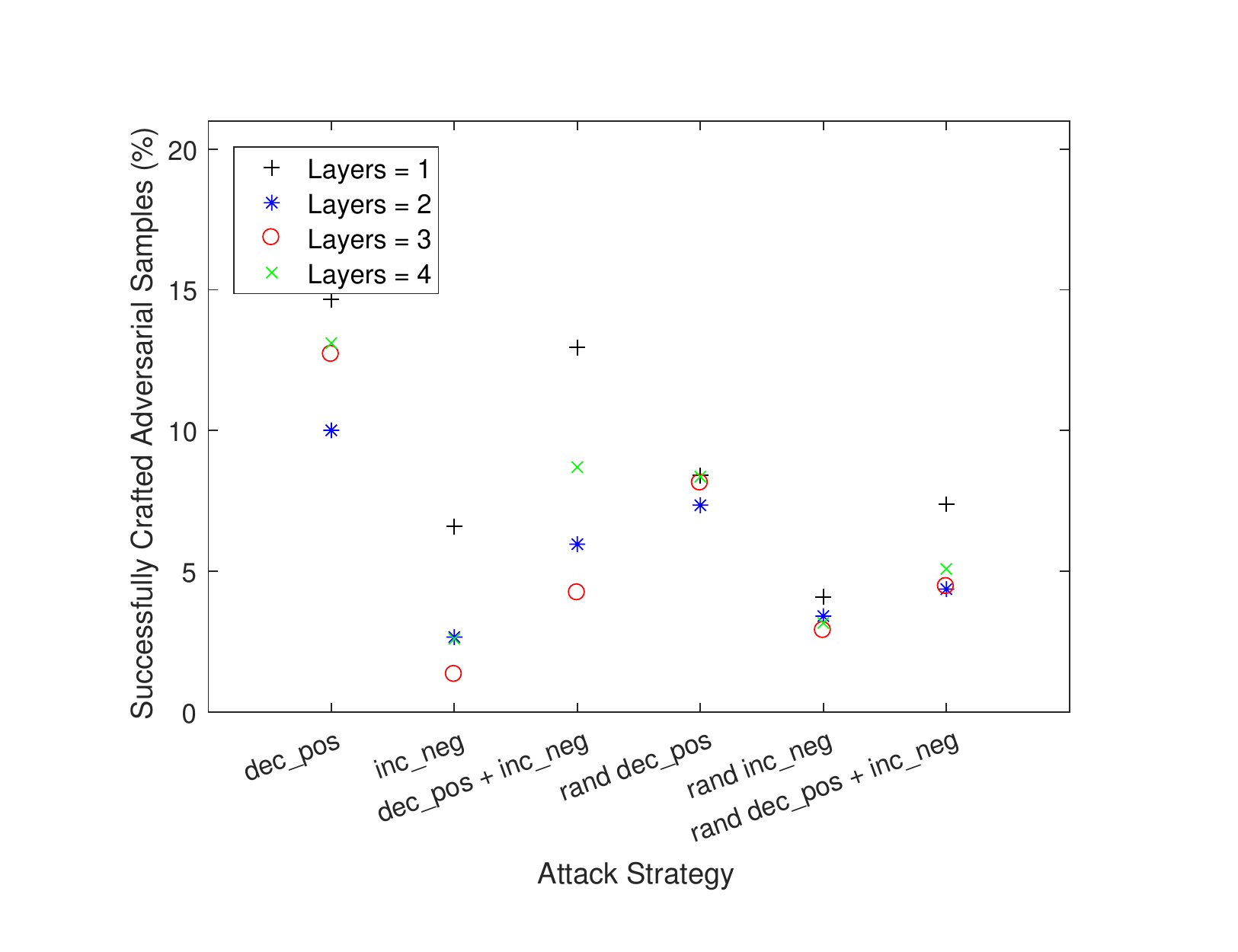}}
\caption{Percentage of successfully crafted adversarial samples for different sample crafting strategies with the distillation defense and $T=10$.}
\label{strategySummaryT10}
\end{figure}

We summarize the success of the different iterative strategies for crafting adversarial samples after iteration 20 in Figures~\ref{strategySummaryT1}
for the baseline classifier, Figure~\ref{strategySummaryT2} for $T = 2$, and Figure~\ref{strategySummaryT10} for $T = 10$.
The figures indicate that shallow networks with $H = 1$ hidden layers are the most susceptible to successfully crafted adversarial samples.
We see that using the Jacobian information can help to craft more adversarial samples with the same number of perturbed features than its randomized counterparts.
From the attacker's perspective,  the dec\_pos strategy (switching off positive malware features) is the most effective approach for crafting adversarial samples for the full defense with $T = 10$.
Likewise, dec\_pos + inc\_neg (alternatively switching off positive feature and switching on negative feature) is more effective than inc\_neg (switching on negative features).
This is fortunate from the defender's perspective because it requires the attacker to potentially spend more effort implementing alternative strategies for
removing malicious features.

%

\begin{figure}[!tbh]
\centering
%\vspace{-0.4in}
{\label{}\includegraphics[trim = 1.25in 3.0in 1.5in 3.0in,clip,width=0.9\columnwidth]{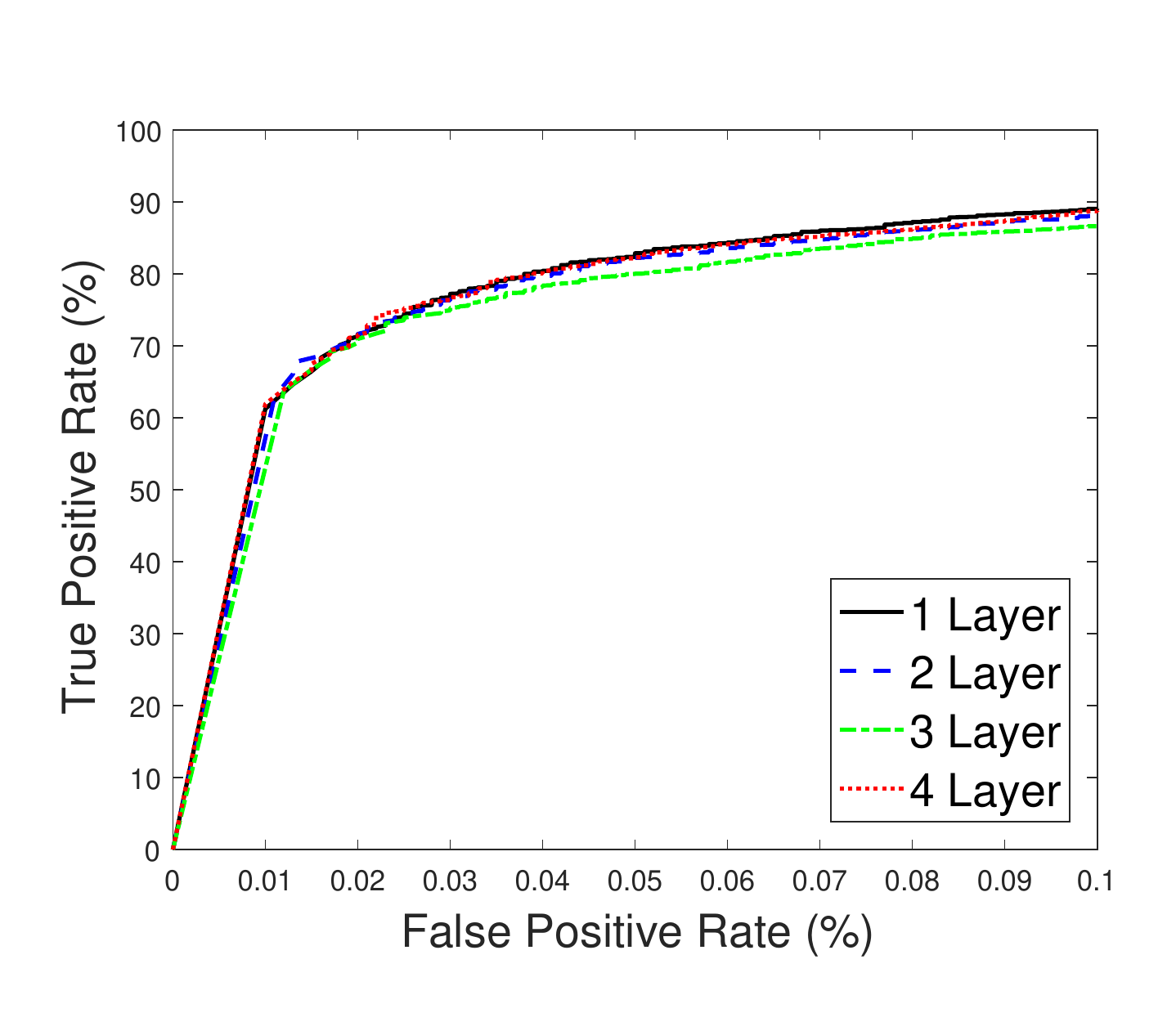}}
\caption{ROC curves of the malware classifiers for the regularization defense with $D=0.0001$ for different numbers of hidden layers.}
\label{fig:Reg-0001-ROC}
\end{figure}

\begin{figure}[!tbh]
\centering
%\vspace{-0.4in}
{\label{}\includegraphics[trim = 1.25in 3.0in 1.5in 3.0in,clip,width=0.9\columnwidth]{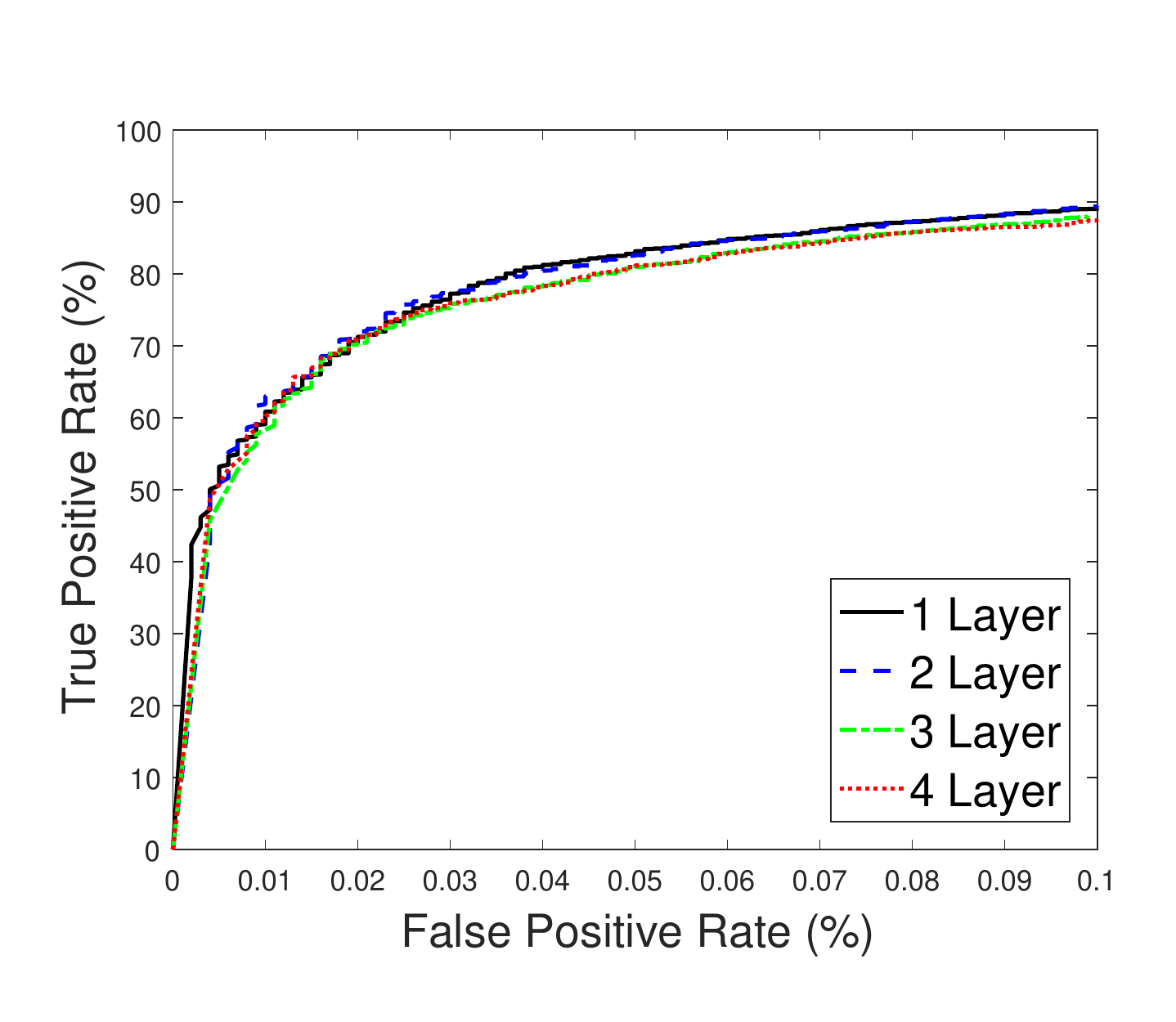}}
\caption{ROC curves of the malware classifiers for the regularization defense with $D=0.0005$ for different numbers of hidden layers.}
\label{fig:Reg-0005-ROC}
\end{figure}

\begin{figure}[!tbh]
\centering
%\vspace{-0.4in}
{\label{}\includegraphics[trim = 1.25in 3.0in 1.5in 3.0in,clip,width=0.9\columnwidth]{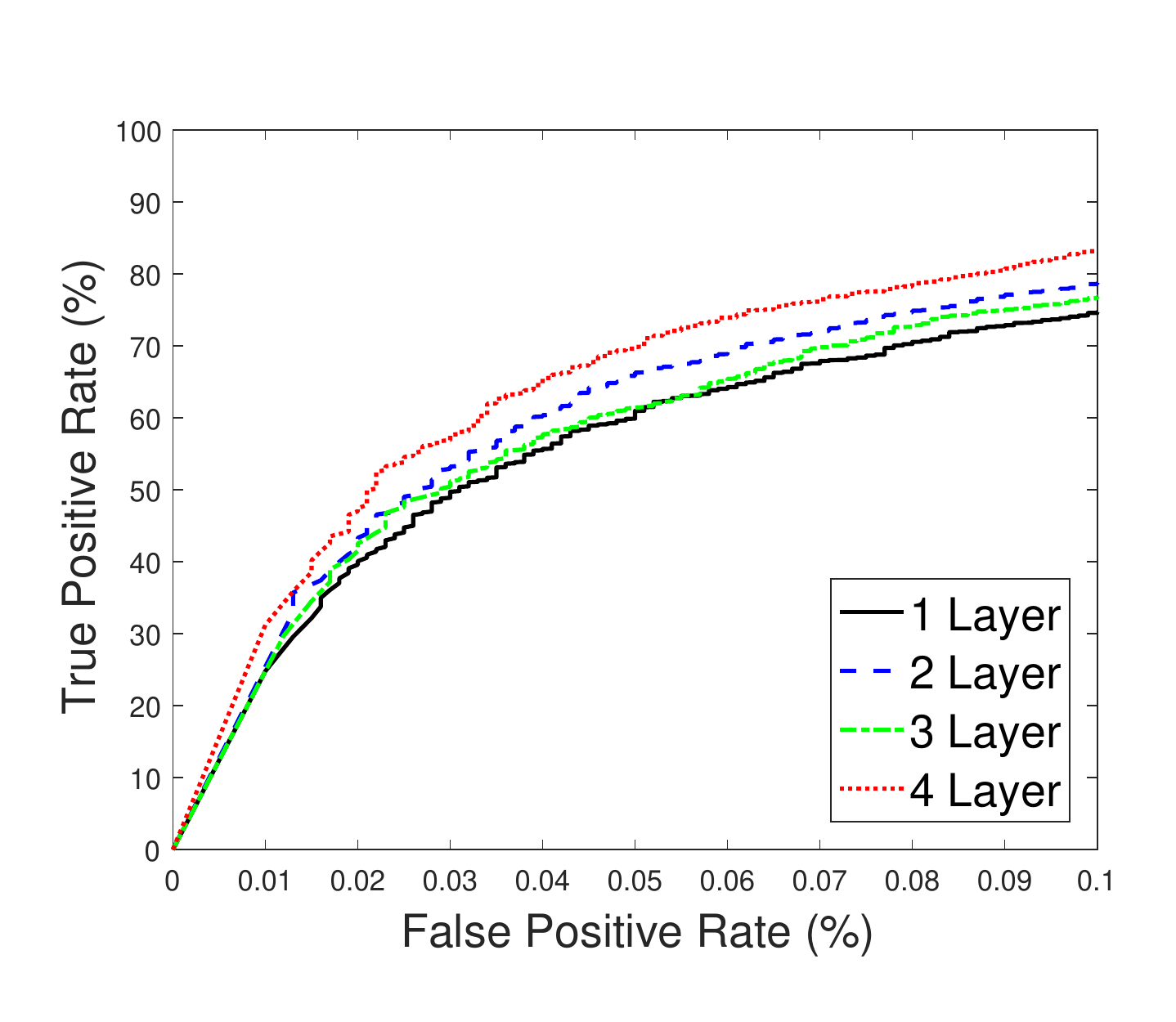}}
\caption{ROC curves of the malware classifiers for the regularization defense with $D=0.001$ for different numbers of hidden layers.}
\label{fig:Reg-001-ROC}
\end{figure}

\begin{figure}[!tbh]
\centering
%\vspace{-0.4in}
{\label{}\includegraphics[trim = 1.25in 3.0in 1.5in 3.0in,clip,width=0.9\columnwidth]{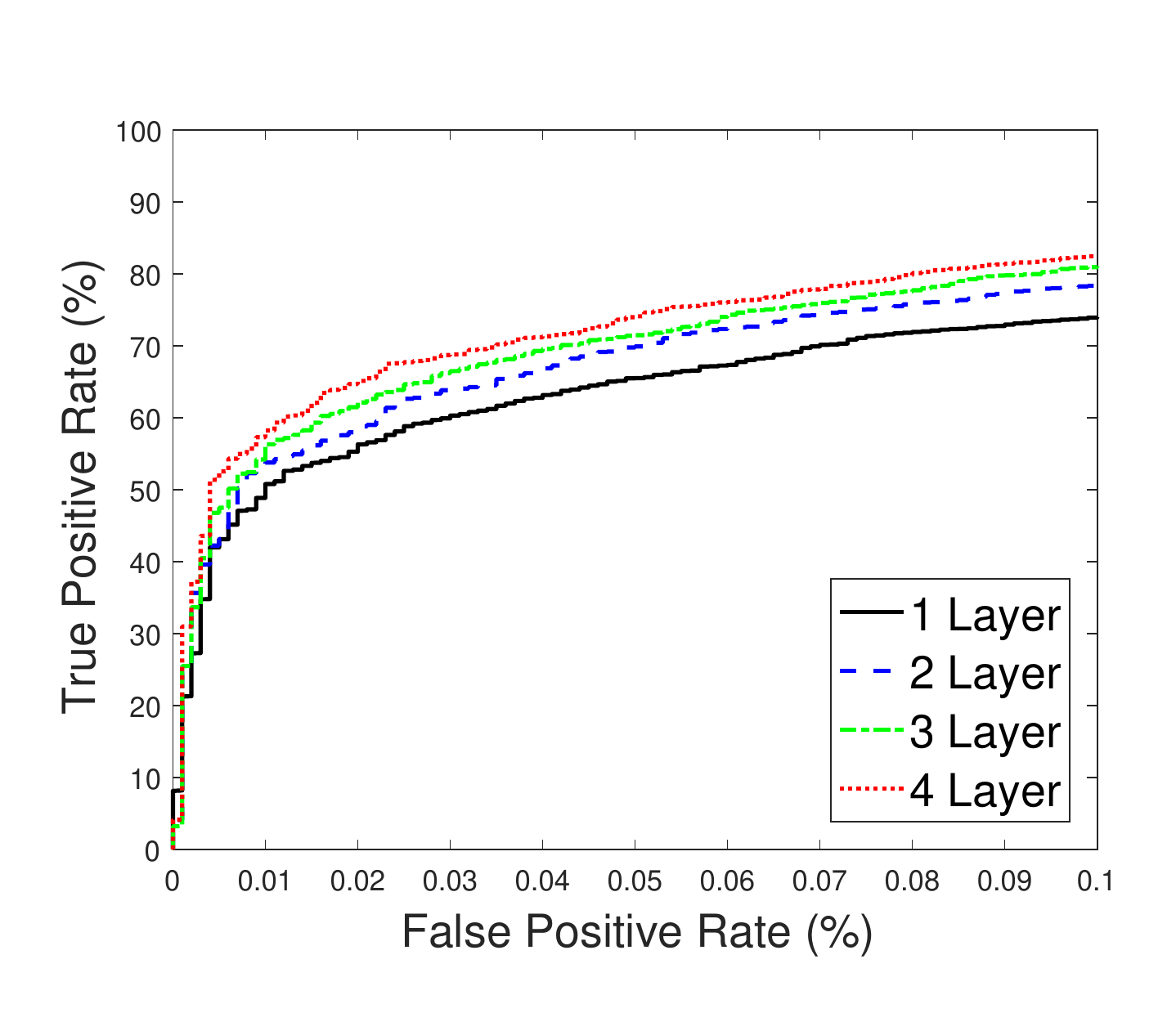}}
\caption{ROC curves of the malware classifiers for the regularization defense with $D=0.01$ for different numbers of hidden layers.}
\label{fig:Reg-01-ROC}
\end{figure}

%\subsection{Weight Decay Defense}
\textbf{Weight Decay Defense:}
We next present an analysis of the proposed weight decay defense. We train the malware classification model using different strengths of
weight decay regularization, $D \in \{0.0001, 0.0005, 0.001, 0.01\}$, and plot the ROC curves for these values of $D$
in Figures~\ref{fig:Reg-0001-ROC}-\ref{fig:Reg-01-ROC}, respectively. In general, the true positive rates drop with increasing values of $D$.

\begin{figure}[!tbh]
\centering
%\vspace{-0.4in}
{\label{}\includegraphics[trim = 1.0in 3.0in 1.0in 3.0in,clip,width=1.0\columnwidth]{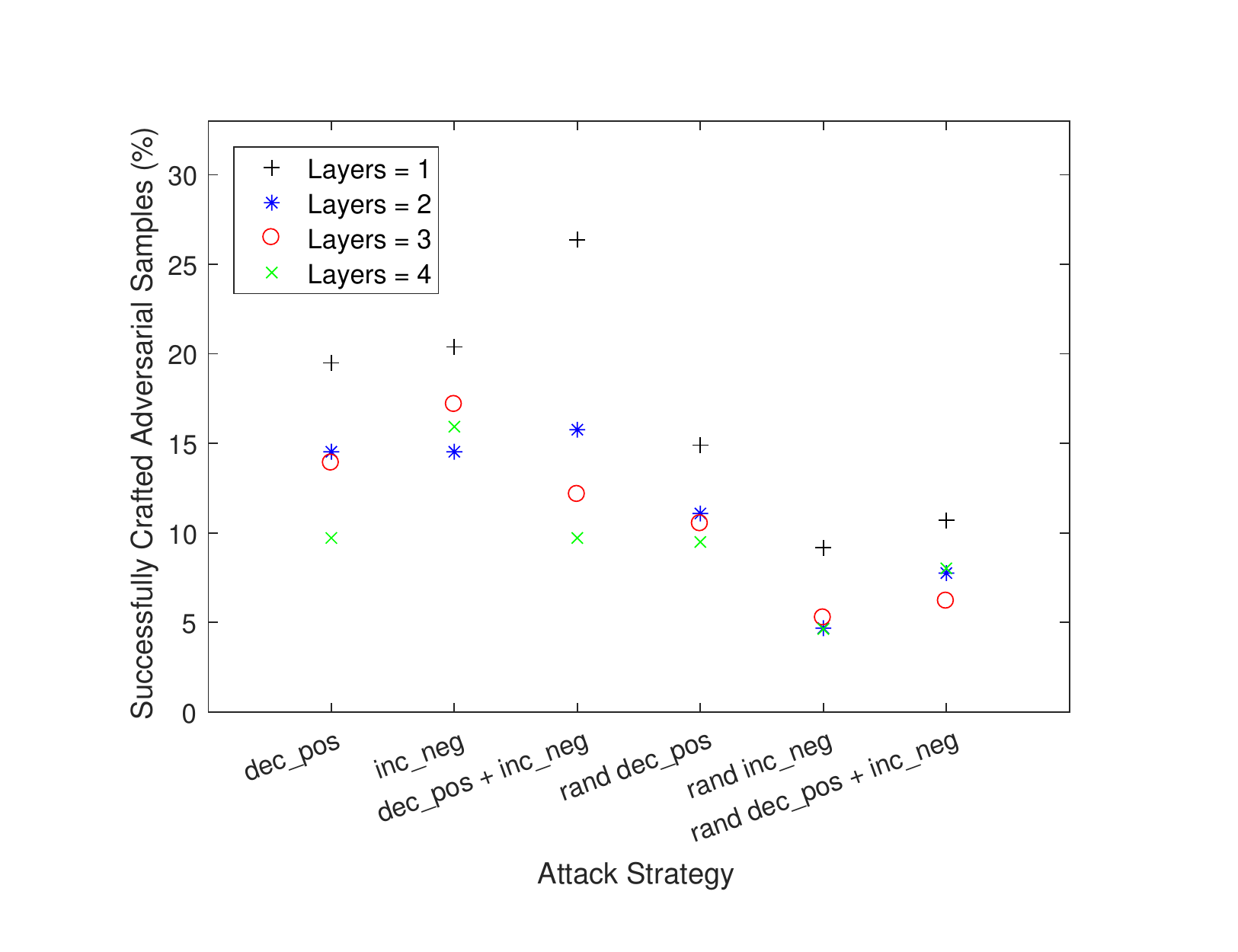}}
\caption{Percentage of successfully crafted adversarial samples after iteration 20 for different sample crafting strategies with the weight decay defense with $D = 0.0001$.}
\label{strategyRegSummaryD0_0001}
\end{figure}

\begin{figure}[!tbh]
\centering
%\vspace{-0.4in}
{\label{}\includegraphics[trim = 1.0in 3.0in 1.0in 3.0in,clip,width=1.0\columnwidth]{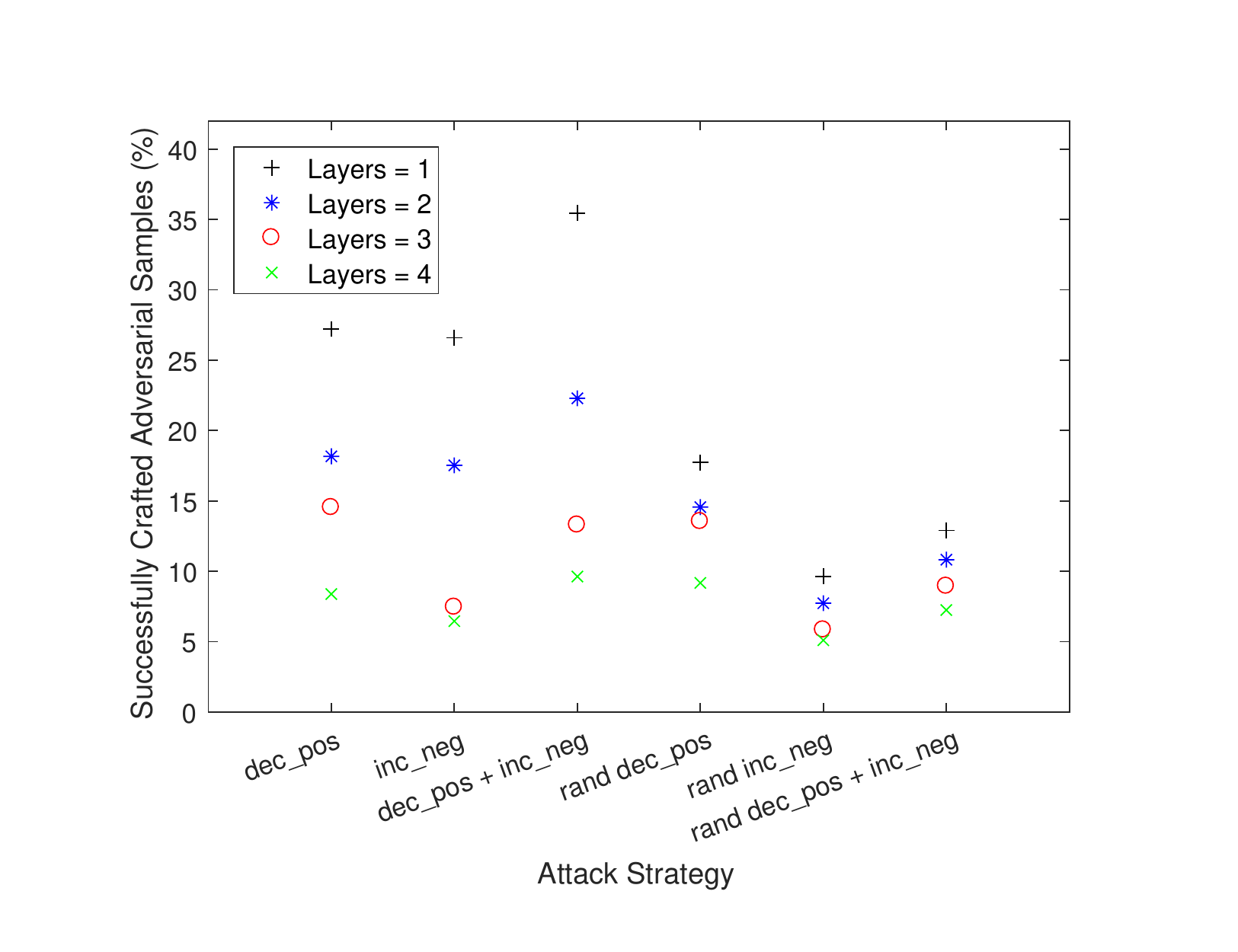}}
\caption{Percentage of successfully crafted adversarial samples after iteration 20 for different sample crafting strategies with the weight decay defense with $D = 0.0005$.}
\label{strategyRegSummaryD0_0005}
\end{figure}

We analyze all combinations of weight decay strength and hidden layer depth in terms of defense to adversarial attacks. The best overall resilience of this model defense to the six adversarial sample
crafting strategies for iteration 20 also employs $D = 0.0001$ and is summarized in Figure~\ref{strategyRegSummaryD0_0001}. For comparison, we also summarize the defensive capabilities
for $D = 0.0005$ in Figure~\ref{strategyRegSummaryD0_0005}.
Figure~\ref{strategyRegSummaryD0_0001} shows that the resilience to adversarial sample crafting strategies also increases as the hidden layer depth increases.
The weight decay defense is not as effective as the distillation defense in Figure~\ref{strategySummaryT10} or even the basline model in Figure~\ref{strategySummaryT1}.

%\begin{comment}
\begin{figure}[!tbh]
\centering
\hspace{0in}
{\label{}\includegraphics[trim = 1.0in 3.0in 1.0in 3.0in,clip,width=1.0\columnwidth]{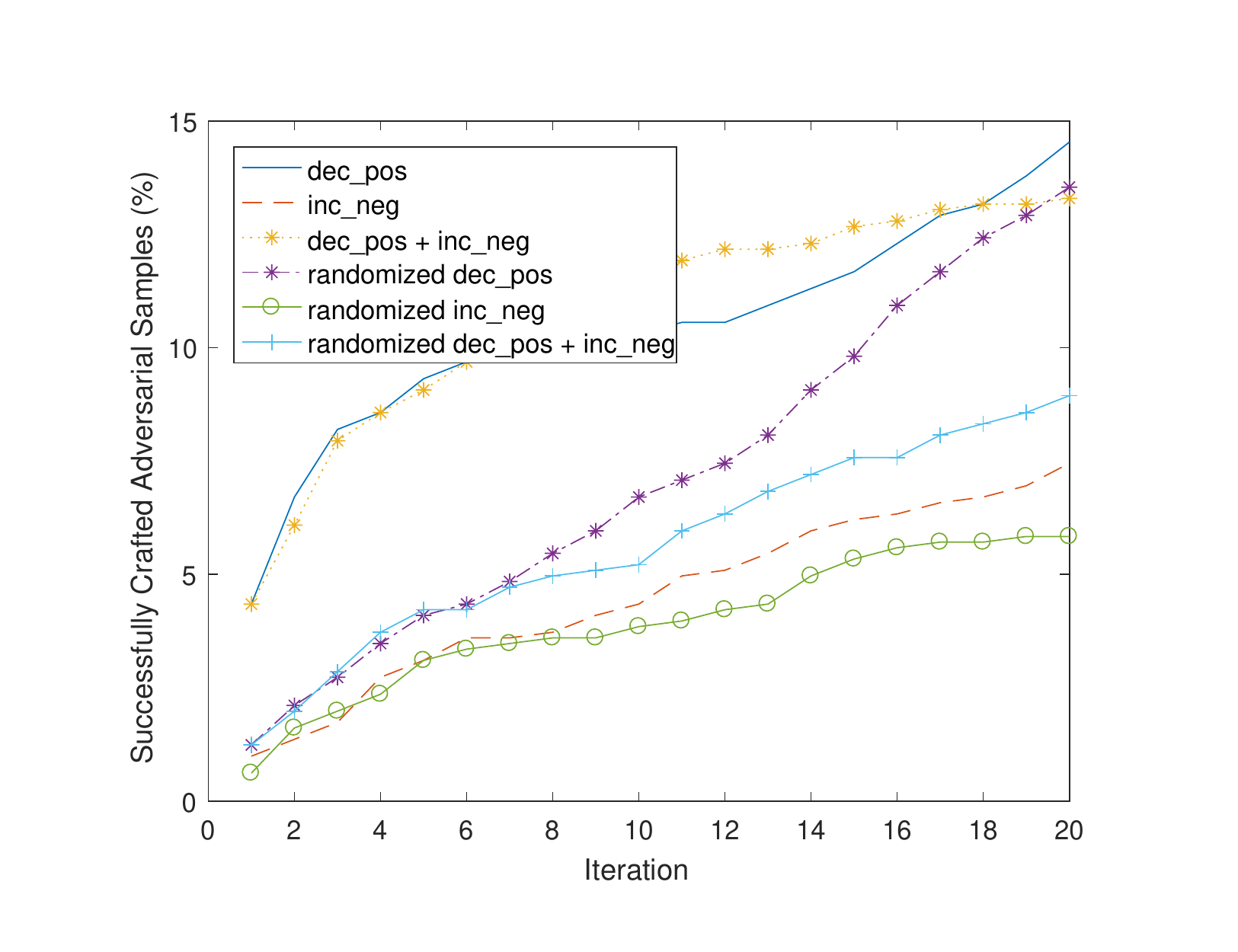}}
\caption{Percentage of successfully crafted adversarial samples for the first 20 iterations with different sample crafting strategies, $D = 0.0005$ weight decay and $L = 3$ hidden layers.}
\label{strategyReg0_0005}
\end{figure}
%\end{comment}

\begin{figure}[!tbh]
\centering
%\vspace{-0.4in}
{\label{}\includegraphics[trim = 1.25in 3.0in 1.5in 3.0in,clip,width=0.9\columnwidth]{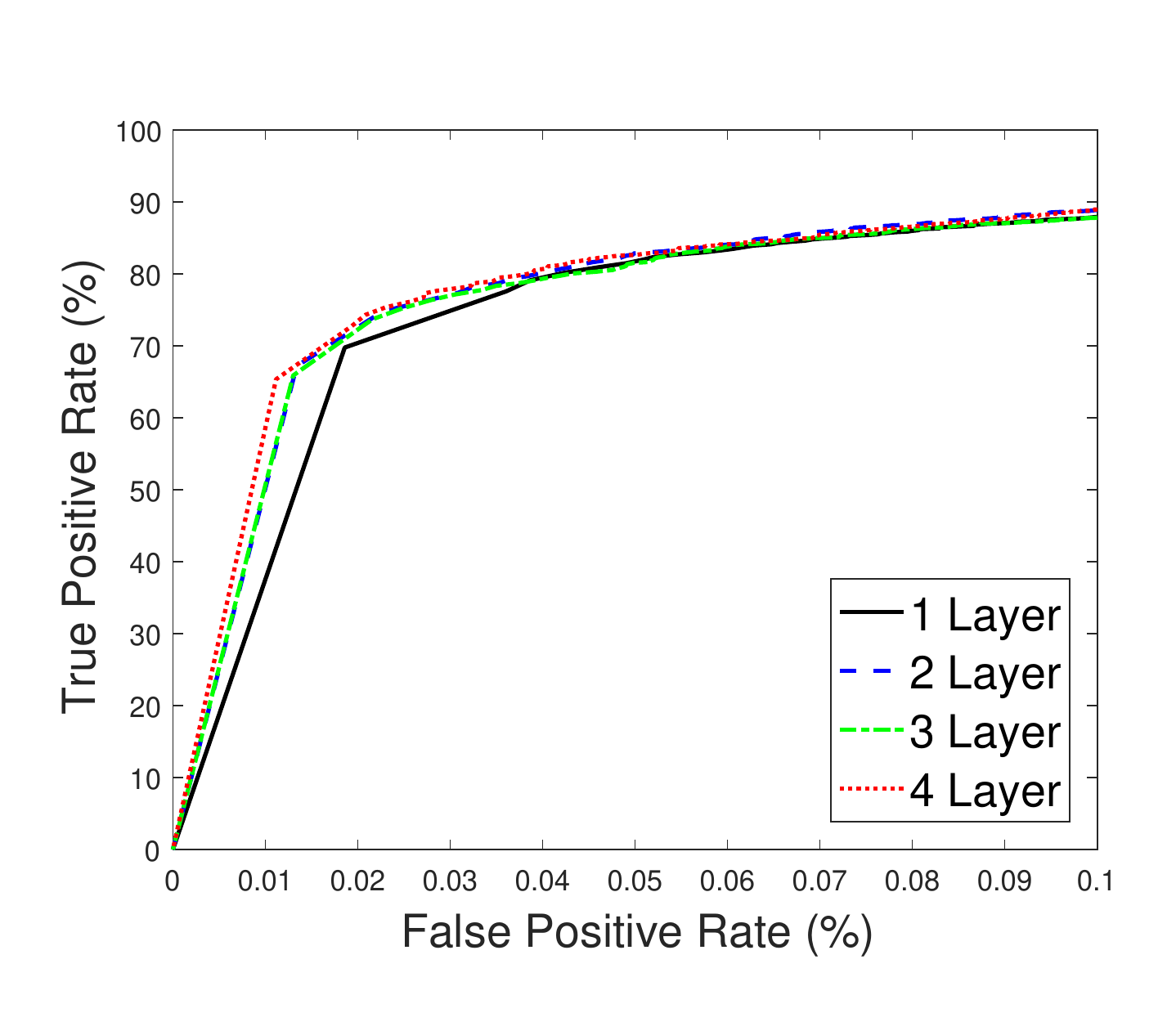}}
\caption{ROC curves of the ensemble malware classifier with $E = 5$ classifiers for different numbers of hidden layers.}
\label{fig:e_5-ROC}
\end{figure}

%\subsection{Ensemble Defense}
\textbf{Ensemble Defense:}
Finally, we present the results for the ensemble defense on our dataset. In Figure~\ref{fig:e_5-ROC}, we present the ROC curves for an ensemble with $E = 5$ classifiers. Ensembles with
other numbers of classifiers offer similar results.

\begin{figure}[!tbh]
\centering
%\vspace{-0.4in}
{\label{}\includegraphics[trim = 1.0in 3.0in 1.0in 3.0in,clip,width=1.0\columnwidth]{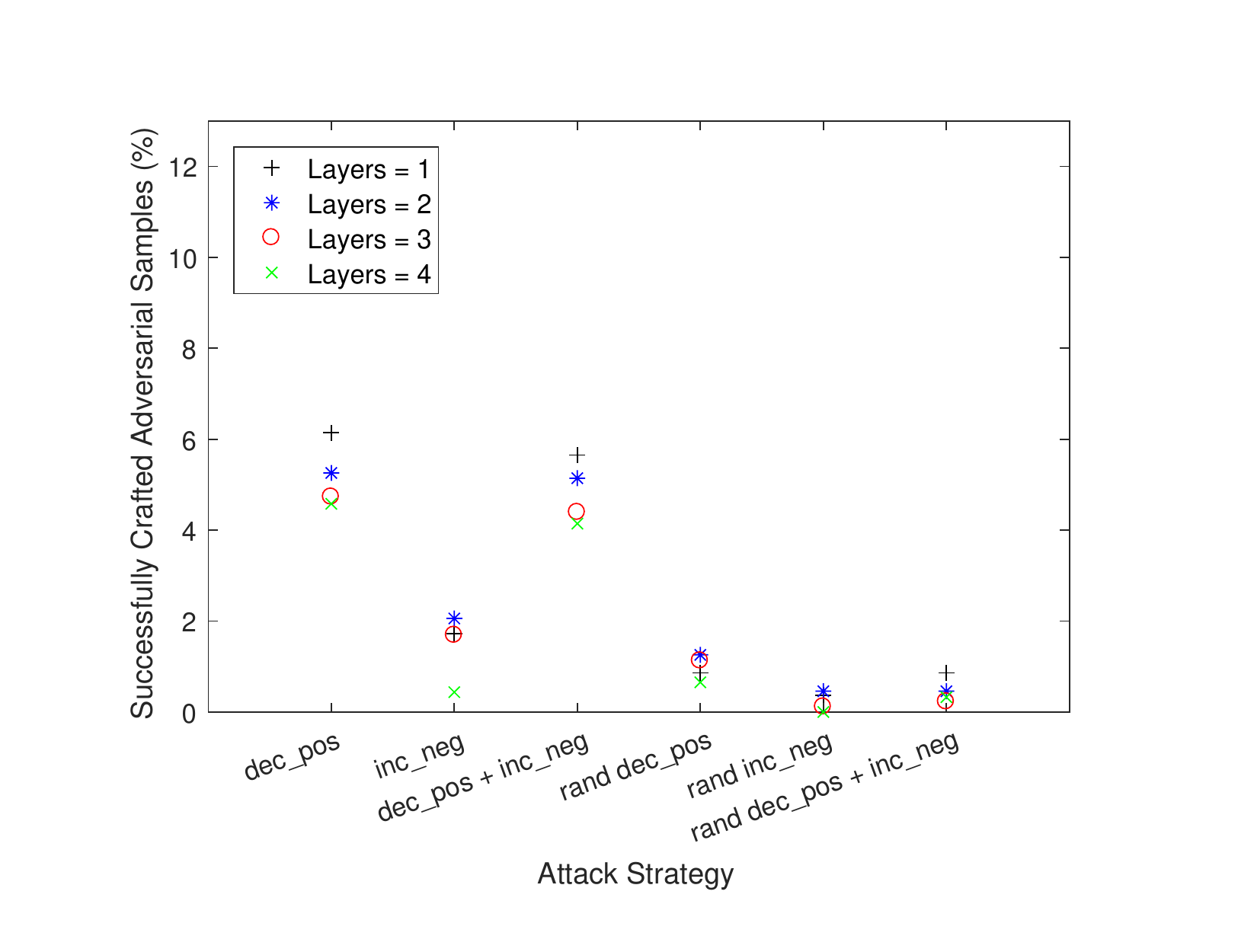}}
\caption{Percentage of successfully crafted adversarial samples after iteration 20 for different sample crafting strategies with the $N = 3$ ensemble defense.}
\label{strategySummaryE3}
\end{figure}

\begin{figure}[!tbh]
\centering
%\vspace{-0.4in}
{\label{}\includegraphics[trim = 1.0in 3.0in 1.0in 3.0in,clip,width=1.0\columnwidth]{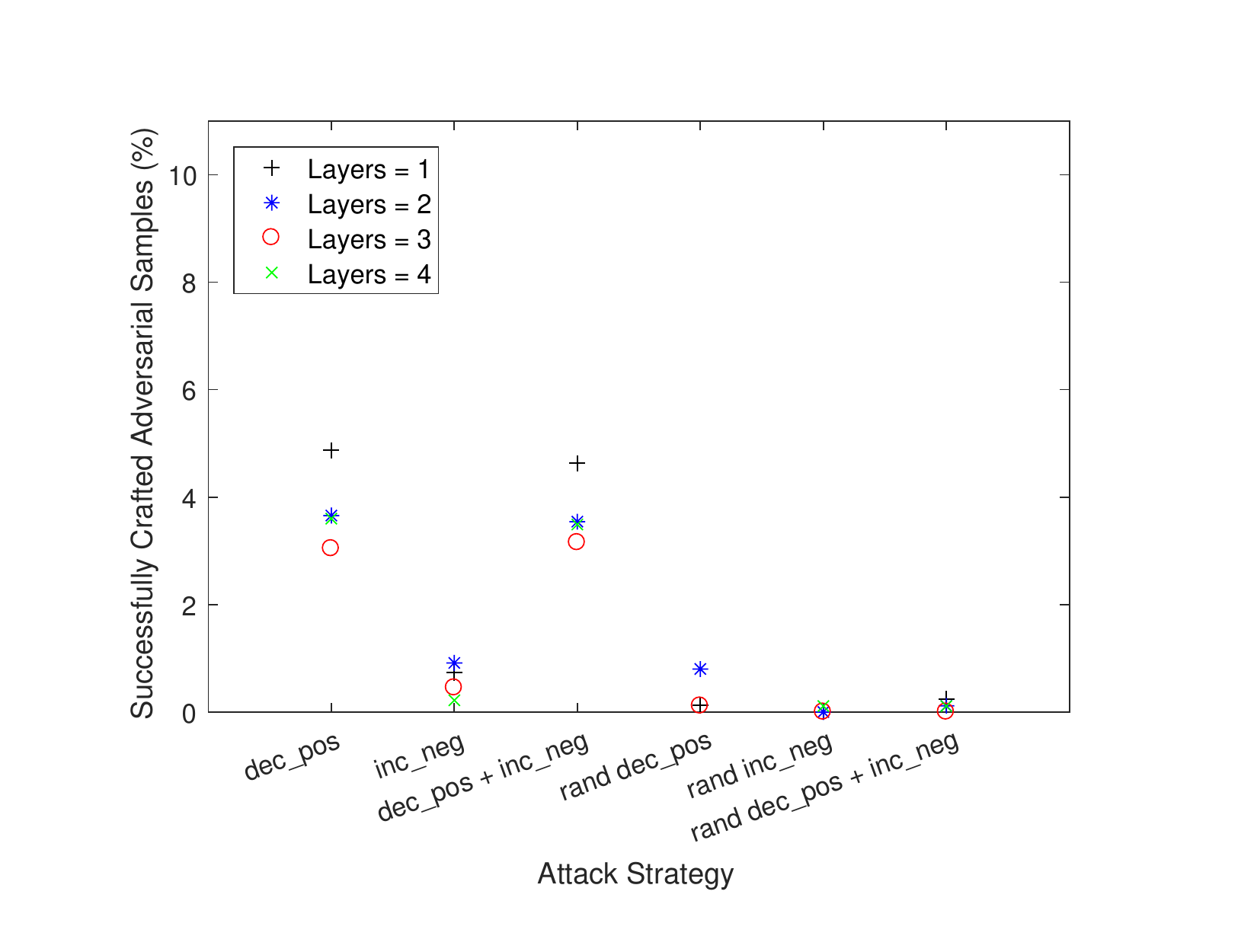}}
\caption{Percentage of successfully crafted adversarial samples after iteration 20 for different sample crafting strategies with the $N = 5$ ensemble defense.}
\label{strategySummaryE5}
\end{figure}

The summary results after 20 iterations for $E = 3$ and $E = 5$ classifiers are shown in Figure~\ref{strategySummaryE3} and Figure~\ref{strategySummaryE5}, respectively.
The figures indicate that increasing the number of classifiers in the ensemble make increases the difficulty of successfully crafting adversarial examples.
Furthermore,
the ensemble defense greatly reduces the percentage of successfully crafted samples compared to the
results for the baseline classifier in Figure~\ref{strategySummaryT1} and the distillation defense with $T = 10$ in Figure~\ref{strategySummaryT10}.

\section{Related Work}
We next present previous research related to our study. First we describe previous papers related to adversarial learning in general. We then discuss papers related to adversarial learning to either create or detect adversarial malware samples.

\textbf{Adversarial Attacks:}
%\subsection{Adversarial Attacks}
Adversarial attacks and defense for deep learning models have been a popular research topic recently due to the wide range applications of deep learning models.
Goodfellow, \emph{et al.},~\cite{goodfellow2014explaining} demonstrated that deep learning models can be fooled by crafting adversarial samples from the original input data by adding a perturbation
on the direction of the sign of the model's cost function gradient. This method is known as the \emph{fast gradient sign} method. For images which are considered
in their paper, the algorithm computes the gradient information once and perturbs all of the pixels to a certain amplitude. Since the fast gradient sign method requires continuous features,
it is not applicable to our malware classification data which is composed of sparse binary features.

Papernot, \emph{et al.},~\cite{papernot2015limitations} proposed another algorithm for crafting adversarial samples, which iteratively perturbs the input along the dimension with largest gradient saliency. The algorithm perturbs one input feature in each iteration until the altered sample is misclassified into the desired target class. The goal of this method is to use the minimum perturbation to the original sample such that the perturbation is not perceivable by humans, but is misclassified by a machine learning model. This algorithm has a larger computational complexity compared to the fast gradient sign method in~\cite{goodfellow2014explaining}, because in each iteration, the algorithm needs to compute the derivative of the model's output probability with respect to the perturbed sample.

In most cases, users do not have the knowledge about the architecture and parameters of the trained model that are deployed into a service. Deployed models are known as black box models due to the fact that the attacker does not have any information beyond the outputs of the model on input queries. Papernot, \emph{et al.},~\cite{papernot2016practical} proposed a method based on model distillation to craft adversarial attack samples on black box models.
The authors in~\cite{papernot2016practical} found that adversarial samples are transferable among models, \emph{i.e.}, the adversarial samples crafted for one model can also mislead the classification of other models. They use model distillation techniques to compromise an oracle hosted by MetaMind. In this case, the oracle is a defensive system where the users only know the input and output, but they do not know anything about the architecture of the model.

A defensive strategy using model distillation is proposed in~\cite{papernot2015distillation}. Model distillation is performed by using the soft labels (prediction probability on a trained neural network) as the label of training samples to train a new deep neural network.  They found that distillation captures class correlation, and the model trained on soft labels is more robust than one trained using hard labels. In this case, a hard label is specified as the discrete class label.
The authors also found that using a high temperature in distillation training enforces smoothness of the model, which could make the model more robust to adversarial samples. Using the high temperature distilled model, the changes in adversarial samples have much less impact on the classification of the model.

Several authors~\cite{Kantchelian2016,Tramer2016,Feng2016,Tramer2017} have proposed using an ensemble of models to avoid different type of malicious attacks.
For example, the authors in~\cite{Tramer2016} proposed using an ensemble of models to improve the privacy of deployed models since
attackers will only be able to obtain an approximation of the target prediction function.
Kantchelian, \emph{et al.},~\cite{Kantchelian2016} proposed two algorithms for evasion attacks on tree ensemble classifiers, like gradient boosted
trees and random forests. However, each tree classifier is very weak compared with a full-fledged neural network.

\textbf{Malware Classification:}
%\subsection{Malware Classification}
%
Several deep learning malware classifiers are proposed in~\cite{dahl2013large,saxe2015,pascanu2015malware,Huang2016,BenMalware,Kolosnjaji}.
The first study of deep learning for a DNN malware classifier was presented in~\cite{dahl2013large}. Similar to our results, the
authors found that a shallow neural network slightly outperformed a DNN on dynamic analysis-based malware classification.
Saxe, \emph{et al.}, studied DNNs in the context of static malware classification in~\cite{saxe2015}. Huang and Stokes proposed a deep, multi-task approach
for dynamic analysis which simultaneously tries to optimize predicting a) if a file is malicious or benign and b) the file's family if it is malware or returning a benign label
in the case it is clean. In~\cite{pascanu2015malware}, the authors propose a two-stage approach where the first stage employs a language-model, using a
recurrent neural network (RNN) or an echo state network (ESN),
to first learn an embedding of the behavior of the file based on its system call events. This embedding then serves as the features for a DNN in the second stage.
Athiwaratkun, \emph{et al.},~\cite{BenMalware} explored similar architectures for deep malware classification using long short-term memory (LSTM) or a gated recurrent units (GRU) for the language model, as well as a separate
architecture using a
character-level convoluation neural network (CNN). In~\cite{Kolosnjaji}, Kolosnjaji, \emph{et al.}, propose an alternative model also employing a CNN and an LSTM.

Several authors have proposed methods for creating adversarial malware samples.
In~\cite{weilinxu2016evading}, Xu, \emph{et al.}, propose a system which uses a genetic algorithm to generate adversarial samples which can be mispredicted by a classifier.
The system assumes access to the classifier's output score. The authors demonstrate that their system can automatically create 500 malicious PDF files
that are classified as benign by the PDFrate~\cite{smutz2012} and Hidost~\cite{Srndic2013} systems.

Hu and Tan~\cite{HuAdversarialMalwareGan} propose a generative adversarial network (GAN) to create adversarial malware samples. In their work, the authors assume that the attackers
know the features which are employed by the malware classifier, but they do not know the classification model or its parameters. They use static analysis where the features
are API calls and a sparse binary feature is constructed to indicate which APIs were called by the program. Furthermore, the authors assume that the prediction score
from the model is reported from the malware classification model.

Grosse, \emph{et al.},~\cite{Grosse2017a} study the distillation defense for \textit{static} analysis-based malware classification. Similar to this paper, the authors assume that the attacker has
access to all of the deep learning malware classifier's model parameter. In our work, we also consider the distillation defense for
\textit{dynamic} analysis-based malware classification.  In addition, we evaluate the ensemble defense and introduce the regularization defense for a dynamic malware classifier.
In another recent paper, Grosse, \emph{et al.},~\cite{Grosse2017b} add a separate class for adversarial samples and propose a statistical hypothesis test to identify adversarial samples.

\section{Conclusion}
In this paper, we investigated six different adversarial learning attack strategies against a dynamic analysis-based, deep learning malware classification
system. We analyzed the effectiveness of two previously proposed  defensive methods including the distillation defense and ensemble defense.
We also proposed and analyzed the weight decay defense. All three defenses offer comparable classification
accuracies compared to a standard deep learning baseline system. Thus, they achieve a key goal in adversarial learning of not significantly reducing
the accuracy compared to a system without any adversarial learning defenses. In addition, \textit{deep} learning models offer better
resilience to adversarial attacks
than the shallow baseline models in all cases.

Results show that the ensemble classifier provides significantly better resiliency against adversarial attacks for this dataset when compared to the other defenses, but requires more computational resources for both training
and inference. The distillation offers the second best resistance, and
helps to reduce the effectiveness of removing important malicious features.
The weight decay defense offers little defense against crafted adversarial samples.




\bibliographystyle{IEEEtran}
\bibliography{security,deep,deep_new}

\begin{thebibliography}{10}
\providecommand{\url}[1]{#1}
\csname url@samestyle\endcsname
\providecommand{\newblock}{\relax}
\providecommand{\bibinfo}[2]{#2}
\providecommand{\BIBentrySTDinterwordspacing}{\spaceskip=0pt\relax}
\providecommand{\BIBentryALTinterwordstretchfactor}{4}
\providecommand{\BIBentryALTinterwordspacing}{\spaceskip=\fontdimen2\font plus
\BIBentryALTinterwordstretchfactor\fontdimen3\font minus
  \fontdimen4\font\relax}
\providecommand{\BIBforeignlanguage}[2]{{%
\expandafter\ifx\csname l@#1\endcsname\relax
\typeout{** WARNING: IEEEtran.bst: No hyphenation pattern has been}%
\typeout{** loaded for the language `#1'. Using the pattern for}%
\typeout{** the default language instead.}%
\else
\language=\csname l@#1\endcsname
\fi
#2}}
\providecommand{\BIBdecl}{\relax}
\BIBdecl

\bibitem{dahl2013large}
G.~E. Dahl, J.~W. Stokes, L.~Deng, and D.~Yu, ``Large-scale malware
  classification using random projections and neural networks,'' in
  \emph{Proceedings of the IEEE International Conference on Acoustics, Speech
  and Signal Processing (ICASSP)}.\hskip 1em plus 0.5em minus 0.4em\relax IEEE,
  2013, pp. 3422--3426.

\bibitem{saxe2015}
J.~Saxe and K.~Berlin, ``Deep neural network based malware detection using
  two-dimensional binary program features,'' \emph{Malware Conference
  (MALCON)}, 2015.

\bibitem{pascanu2015malware}
R.~Pascanu, J.~W. Stokes, H.~Sanossian, M.~Marinescu, and A.~Thomas, ``Malware
  classification with recurrent networks,'' in \emph{Proceedings of the IEEE
  International Conference on Acoustics, Speech and Signal Processing
  (ICASSP)}.\hskip 1em plus 0.5em minus 0.4em\relax IEEE, 2015, pp. 1916--1920.

\bibitem{Huang2016}
W.~Huang and J.~W. Stokes, ``Mtnet: A multi-task neural network for dynamic
  malware classfication,'' in \emph{Proceedings of Detection of Intrusions and
  Malware, and Vulnerability Assessment (DIMVA)}, 2016, pp. 399--418.

\bibitem{BenMalware}
B.~Athiwaratkun and J.~W. Stokes, ``Malware classification with lstm and gru
  language models and a character-level cnn,'' in \emph{2017 IEEE International
  Conference on Acoustics, Speech and Signal Processing (ICASSP)}, March 2017,
  pp. 2482--2486.

\bibitem{Kolosnjaji}
B.~Kolosnjaji, A.~Zarras, G.~Webster, and C.~Eckert, ``Deep learning for
  classification of malware system call sequences,'' in \emph{Australasian
  Joint Conference on Artificial Intelligence}.\hskip 1em plus 0.5em minus
  0.4em\relax Springer International Publishing, 2016, pp. 137--149.

\bibitem{papernot2016practical}
N.~Papernot, P.~McDaniel, I.~Goodfellow, S.~Jha, Z.~B. Celik, and A.~Swami,
  ``Practical black-box attacks against deep learning systems using adversarial
  examples,'' \emph{Proceedings of the ACM Asia Conference on Computer and
  Communications Security}, 2017.

\bibitem{goodfellow2014explaining}
I.~J. Goodfellow, J.~Shlens, and C.~Szegedy, ``Explaining and harnessing
  adversarial examples,'' \emph{Proceedings of the International Conference on
  Learning Representations (ICML)}, 2015.

\bibitem{nguyen2015deep}
A.~Nguyen, J.~Yosinski, and J.~Clune, ``Deep neural networks are easily fooled:
  High confidence predictions for unrecognizable images,'' in \emph{Proceedings
  of the IEEE Conference on Computer Vision and Pattern Recognition (CVPR)},
  2015, pp. 427--436.

\bibitem{weilinxu2016evading}
W.~Xu, Y.~Qi, and D.~Evans, ``Automatically evading classifiers,''
  \emph{Proceedings of the Network and Distributed System Security Symposium
  (NDSS)}, 2016.

\bibitem{papernot2015distillation}
N.~Papernot, P.~McDaniel, X.~Wu, S.~Jha, and A.~Swami, ``Distillation as a
  defense to adversarial perturbations against deep neural networks,''
  \emph{Proceedings of the 37th IEEE Symposium on Security and Privacy}, 2015.

\bibitem{Kantchelian2016}
A.~Kantchelian, J.D.Tygar, and A.~D.Joseph, ``Evasion and hardening of tree
  ensemble classifiers,'' \emph{Proceedings of the International Conference on
  Machine Learning (ICML)}, 2016.

\bibitem{Tramer2016}
F.~Tramer, F.~Zhang, A.~Juels, M.~K. Reiter, and T.~Ristenpart, ``Stealing
  machine learning models via prediction apis,'' \emph{Proceedings of the
  USENIX Security Symposium}, 2016.

\bibitem{Feng2016}
J.~Feng, T.~Zahavy, B.~Kang, H.~Xu, and S.~Mannor, ``Ensemble robustness of
  deep learning algorithms,'' \emph{arXiv preprint arXiv:1602.02389v3}, 2016.

\bibitem{Tramer2017}
F.~Tramer, A.~Kurakin, N.~Papernot, D.~Boneh, and P.~McDaniel, ``Ensemble
  adversarial training: Attacks and defense,'' \emph{arXiv preprint
  arXiv:1705.07204v2}, 2017.

\bibitem{Xu2017}
W.~Xu, D.~Evans, and Y.~Qi, ``Feature squeezing: Detecting adversarial examples
  in deep neural networks,'' \emph{arXiv preprint arXiv:1704.01155v1}, 2017.

\bibitem{Tong2017}
L.~Tong, B.~Li, C.~Hajaj, C.~Xiao, and Y.~Vorobeychik, ``Hardening classifiers
  against evasion: the good, the bad, and the ugly,'' \emph{arXiv preprint
  arXiv:1708.08327v2}, 2017.

\bibitem{HuAdversarialMalwareGan}
W.~Hu and Y.~Tan, ``Generating adversarial malware examples for black-box
  attacks based on gan,'' \emph{arXiv preprint 1702.05983}, 2017.

\bibitem{Grosse2017a}
K.~Grosse, N.~Papernot, P.~Manoharan, M.~Backes, and P.~McDaniel, ``Adversarial
  perturbations against deep neural networks for malware classification,'' in
  \emph{Proceedings of the European Symposium on Research in Computer Security
  (ESORICS)}, 2017.

\bibitem{Grosse2017b}
K.~Grosse, P.~Manoharan†, N.~Papernot, M.~Backes, and P.~McDaniel, ``On the
  (statistical) detection of adversarial examples,'' in \emph{arXiv preprint
  arXiv:1702.06280v2}, 2017.

\bibitem{Manning09}
C.~D. Manning, P.~Raghavan, and H.~Schutze, \emph{An Introduction to
  Information Retrieval}.\hskip 1em plus 0.5em minus 0.4em\relax Cambridge
  University Press, 2009.

\bibitem{papernot2015limitations}
N.~Papernot, P.~McDaniel, S.~Jha, M.~Fredrikson, Z.~B. Celik, and A.~Swami,
  ``The limitations of deep learning in adversarial settings,''
  \emph{Proceedings of the 1st IEEE European Symposium on Security and
  Privacy}, 2015.

\bibitem{Crandall05}
J.~Crandall, Z.~Su, F.~Chong, and S.~Wu, ``On deriving unknown vulnerabilities
  from zero-day polymorphic and metamorphic worm exploits,'' in
  \emph{Proceedings of the ACM Conference on Computer and Communications
  Security (CCS'05)}, 2005, pp. 235--248.

\bibitem{li2006}
P.~Li, T.~J. Hastie, and K.~W. Church, ``Very sparse random projections,'' in
  \emph{Proceedings of the ACM SIGKDD International Conference on Knowledge
  Discovery and Data Mining (ICDM)}, 2006, pp. 287--296.

\bibitem{Srivastava:2014:DSW:2627435.2670313}
\BIBentryALTinterwordspacing
N.~Srivastava, G.~Hinton, A.~Krizhevsky, I.~Sutskever, and R.~Salakhutdinov,
  ``Dropout: A simple way to prevent neural networks from overfitting,''
  \emph{J. Mach. Learn. Res.}, vol.~15, no.~1, pp. 1929--1958, Jan. 2014.
  [Online]. Available: \url{http://dl.acm.org/citation.cfm?id=2627435.2670313}
\BIBentrySTDinterwordspacing

\bibitem{hinton2015distilling}
G.~Hinton, O.~Vinyals, and J.~Dean, ``Distilling the knowledge in a neural
  network,'' \emph{Neural Information Processing Systems (NIPS) Deep Learning
  Workshop}, 2014.

\bibitem{CNTK}
F.~Seide and A.~Agarwal, ``Cntk: Microsoft's open-source deep-learning
  toolkit,'' in \emph{Proceedings of the 22nd ACM SIGKDD International
  Conference on Knowledge Discovery and Data Mining}.\hskip 1em plus 0.5em
  minus 0.4em\relax ACM, 2016, pp. 2135--2135.

\bibitem{smutz2012}
C.~Smutz and A.~Stavrou, ``Malicious pdf detection using metadata and
  structural features,'' \emph{Technical report}, 2012.

\bibitem{Srndic2013}
N.~Srndic and P.~Laskov, ``Detection of malicious pdf files based on
  hierarchical document structure,'' in \emph{Proceedings of the Network and
  Distributed System Security Symposium (NDSS)}, 2013.

\end{thebibliography}

\end{document}